\documentclass[aps,prd,nofootinbib,twocolumn,superscriptaddress,letterpaper,preprintnumbers]{revtex4-1}
\usepackage{amsmath, amssymb,braket, slashed, float, graphicx, subfigure, natbib, mathtools, color} % for details on the use of the package, please
\usepackage[T1]{fontenc} % if needed
\usepackage[colorlinks=true
,urlcolor=blue
,anchorcolor=blue
,citecolor=blue
,filecolor=blue
,linkcolor=red
,menucolor=blue
,hyperfootnotes=false,
,linktocpage=true
,pdfproducer=medialab
,pdfa=true
]{hyperref}
\usepackage[utf8]{inputenc}
\usepackage{xcolor}
\usepackage{cancel, multirow, dsfont}
\usepackage{soul}
\usepackage{cleveref}
\usepackage[compat=1.0.0]{tikz-feynman}

\newcommand{\fnl}{f_{\rm NL}}
\newcommand{\D}{{\rm d}}
\newcommand{\mt}{{\tilde{m}}}
\newcommand{\eftcut}{\Lambda_{\rm EFT}}

\let\vec\mathbf
\newcommand{\es}[2] {\begin{equation} \label{#1} \begin{split} #2 \end{split} \end{equation}}
\newcommand{\ii}{{\rm i}}
\newcommand{\dd}{{\rm d}}
\newcommand{\hc}{\text{h.c.}}
\usepackage{braket}
\usepackage{amssymb}
\usepackage{float}
\usepackage{appendix}
\usepackage{physics}
\usepackage{comment}

\begin{document}
\title{Scalars at the Cosmological Collider: \\ Full Shapes of Tree Diagrams and Bispectrum Searches using {\it Planck} Data
%Cosmological Collider Searches for Heavy Scalars using {\it Planck} Data: \\Tree Diagrams and Full Shapes
}
\author{Soubhik Kumar}
\affiliation{Institute of Cosmology, Department of Physics and Astronomy, Tufts University, Medford, MA 02155, USA}
\author{Qianshu Lu}
\affiliation{School of Natural Sciences, Institute for Advanced Study, Princeton, NJ 08540, USA}
\affiliation{Center for Cosmology and Particle Physics, Department of Physics, New York University, New York, NY 10003, USA}
\author{Zhong-Zhi Xianyu}
\affiliation{Department of Physics, Tsinghua University, Beijing 100084, China}
\affiliation{Peng Huanwu Center for Fundamental Theory, Hefei, Anhui 230026, China }
\author{Yisong Zhang}
\affiliation{Department of Physics, Tsinghua University, Beijing 100084, China}
%\emailAdd{soubhik.kumar@tufts.edu, qianshu.lu@ias.edu}

\begin{abstract}
The Cosmological Collider (CC) provides a unique opportunity to probe the particle spectrum and fundamental interactions at extremely high energies. Massive particles, via their decay into inflaton quanta, can induce a non-analytic, oscillatory, primordial non-Gaussianity (NG), including the bispectrum. At tree level, three classes of such processes contribute to the bispectrum: `single exchange', `double exchange', and `triple exchange', depending on the number of massive particle propagators. We provide a unified evaluation of all three diagrams and derive the explicit shape functions for the bispectrum, valid across the entire kinematic space. We perform a search for these three processes with the {\it Planck} data, finding no evidence for NG. We also consider simple extensions of the minimal scenario that can counter the exponential suppression of the non-analytic signature, and produce {\it on-shell} particles with masses $M\gg H$, the Hubble scale during inflation. In particular, we focus on the `scalar chemical potential' mechanism and extend our previous search to a wider range of chemical potential ($\omega$) and $M$, finding global 1.5$\sigma$ evidence for non-zero NG for the parameter space $\omega - M \simeq 3H$.
\end{abstract}

\maketitle
\tableofcontents

\section{Introduction}

Cosmic inflation presents a unique opportunity to search for new particles through their imprints on the primordial density perturbations that we can observe today in cosmic microwave background (CMB), large-scale structure (LSS), and line intensity mapping surveys. 
The Hubble scale of inflation $H$ could be as high as $4\times 10^{13}$ GeV~\cite{BICEP:2021xfz}, substantially beyond the reach of terrestrial experiments in the near future. 
Heavy particles having masses of order $H$ can be produced during inflation, and from their decays into inflaton quanta, lead to non-Gaussian (NG) correlations of density perturbations.
Thus, the inflationary universe operates as a very high energy ``cosmological collider''. 
This idea has been extensively studied since its inception \cite{Chen:2009we,Chen:2009zp,Chen:2012ge,Pi:2012gf,Gong:2013sma,Arkani-Hamed:2015bza}, and substantial progress has been made both in terms of developing techniques to deal with quantum field theory computations in curved spacetime and exploring mechanisms and models that produce observable signals. 

\begin{figure}[h]
    \centering
\includegraphics[width=0.98\linewidth]{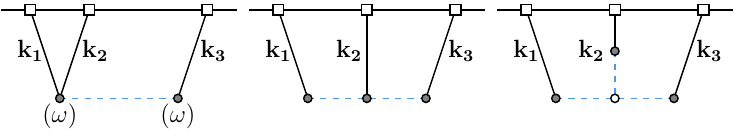}
    \caption{Topologies of `in-in' diagrams considered in this work. The solid black lines are inflaton perturbations $\varphi$ and the dashed blue lines are the massive scalar $\sigma$ propagators. From left to right, we label these as `single exchange' (SE), `double exchange' (DE), and `triple exchange' (TE) topologies, based on the number of $\sigma$ propagators. The $(\omega)$ on the interaction vertices in the SE topology denote the possibility of energy injection in the scalar chemical potential mechanism~\cite{Bodas:2020yho}. Note, for the SE and DE, the full inflaton correlation function $\expval{\varphi_{\vec{k}_1}\varphi_{\vec{k}_2}\varphi_{\vec{k}_3}}$ is a sum over cyclic permutation of momenta assigned to the $\varphi$ legs in the diagram. %\SK{is it possible to make the individual figures a bit larger and may be reduce the horizontal spacing in between?}\ql{I reduced the space and removed the $\eta$ arrow and the made the font bigger.} \SK{great!}
    }
    \label{fig:topo}
\end{figure}

One particular category of cosmological collider scenario that has attracted significant interest~\cite{Baumann:2011su, Assassi:2013gxa, Craig:2014rta,  Dimastrogiovanni:2015pla, Lee:2016vti, Meerburg:2016zdz, Chen:2016uwp, Chen:2016nrs, Chen:2016hrz, An:2017hlx, Chen:2017ryl, Kumar:2017ecc, Baumann:2017jvh,  Chen:2018xck, Kumar:2018jxz, Wu:2018lmx, Dimastrogiovanni:2018uqy, Lu:2019tjj, Hook:2019zxa, Hook:2019vcn, Kumar:2019ebj, Wang:2019gbi, Li:2019ves, Alexander:2019vtb, Bodas:2020yho, Wang:2020ioa, Lu:2021gso, Lu:2021wxu, Dimastrogiovanni:2021cif, Cui:2021iie, Tong:2022cdz, Qin:2022lva, Reece:2022soh, Chen:2022vzh, Maru:2022bhr, Chen:2023txq,   Craig:2024qgy, Quintin:2024boj, Bodas:2024hih, Hubisz:2024xnj, Chakraborty:2025myb, deRham:2025mjh, Bodas:2025vpb,  Chakraborty:2025mhh, Aoki:2025uff, Kumar:2025anx, Jiang:2025mlm} is the on-shell production of scalar particles with masses greater than $3H/2$.
For such masses, the NG induced by these scalars exhibit distinctive oscillatory patterns. 
The simplest such NG is the three-point function (bispectrum) of inflaton fluctuations.
% which we denote by $\sigma$ for a real scalar and $\chi$ for complex.
At tree level, there are only three topologies of Feynman diagrams that contribute to the bispectrum mediated by the heavy scalars: single exchange (SE), double exchange (DE), or triple exchange (TE) of heavy scalars between the external inflaton legs; see Fig.~\ref{fig:topo}.
We will refer to the three topologies collectively as `multiple exchange'. 

In addition to heavy particle production from vacuum fluctuations, which is only efficient for producing particles with mass $M \sim H$, several mechanisms~\cite{Chen:2018xck, Hook:2019zxa, Bodas:2020yho, Chen:2022vzh, Wang:2020ioa} have been discovered in which additional energy sources are present.
These sources can inject energy $\sim \omega $ into the interaction vertices between the inflaton and the heavy scalar, and for $\omega>M\gg H$, produce on-shell particles with $M\gg H$.  
%the interaction vertices between the inflaton and the heavy scalar could provide another source of energy injection $\omega H$, on top of the inflationary expansion which provides $\sim H$ energy.
%This enhances the on-shell production of heavy scalars for mass $M$ satisfying $\omega H>M\gg H$. 
One such mechanism is `scalar chemical potential' (SCP)~\cite{Bodas:2020yho}, where the rolling of the homogeneous inflaton field $\phi$ gives rise to a chemical potential for a complex heavy scalar through a dimension-5 $\partial_{\mu}\phi J_{U(1)}^{\mu}/\Lambda$ coupling, where $J_{U(1)}^{\mu}$ is the $U(1)$ current.\footnote{Another possible mechanism can operate in the presence of primordial features~\cite{Chen:2022vzh}, where classical features on the inflaton potential act as a source of energy injection, transmitted to the heavy scalar through a similar dimension-5 operator. One key difference between the primordial feature and the scalar chemical potential mechanism is that the former breaks scale invariance and modifies the power spectrum. A non-unit speed of sound is also known to enhance the cosmological collider signal~\cite{Pimentel:2022fsc, Jazayeri:2022kjy}, but it is generically generated from a dimension-8 operator in slow-roll inflation. Thus its effect is more suppressed for weakly coupled theories compared to the scalar chemical potential and primordial feature scenarios, where bispectra are generated via dimension-5 operators.}\footnote{Even though in the scalar chemical potential model, the bispectrum also comes from diagrams with a ``single exchange'' topology, by ``single exchange'' we will always refer to scenario where there is no energy injection ($\omega=0$), unless otherwise specified.}
 
In all these scenarios, the on-shell production of the heavy scalar contributes to an oscillatory feature in the bispectrum in the squeezed limit~\cite{Chen:2012ge, Arkani-Hamed:2015bza}, where the momentum of one inflaton fluctuation $\varphi$ is softer than the other two:
\es{eq:ccp_schematic}{
    \frac{\expval{\varphi_{\vec{k_1}}\varphi_{\vec{k_2}}\varphi_{\vec{k_3}}}'}{\expval{\varphi_{\vec{k_1}}\varphi_{-\vec{k_1}}}'\expval{\varphi_{\vec{k_3}}\varphi_{-\vec{k_3}}}'}\xrightarrow{k_3 \ll k_1}\supset e^{-\pi|\nu-\omega/H|}\left(\frac{k_3}{k_1}\right)^{\frac{3}{2}+\ii \nu},
    %\quad \nu \equiv \sqrt{\frac{M^2}{H^2}-\frac{9}{4}},
}
with $\nu = \sqrt{{M^2/H^2}-9/4}>\omega/H$ and $\langle\cdots\rangle = \langle\cdots\rangle'(2\pi)^3\delta(\vec{k}_1+\cdots+\vec{k}_n)$.
For $\omega/H > \nu$, the exponential suppression in eq.~\eqref{eq:ccp_schematic} is absent, since the energy injection counters the exponential Boltzmann suppression to produce the heavy particle on-shell.
% and $\omega$ is the frequency of possible external source of energy injection countering the exponential Boltzmann suppression to produce the heavy $\sigma$ on-shell. 
Notably, the oscillatory feature in eq.~\eqref{eq:ccp_schematic} cannot be mimicked by local inflaton self-interaction, and thus it is a `smoking gun' signature of new particles during inflation.

Despite the strong observational motivation (e.g., see~\cite{Cabass:2024wob,Goldstein:2024bky,Sohn:2024xzd,Suman:2025vuf,Suman:2025tpv,Philcox:2025wts,Bao:2025onc,Anbajagane:2025uro, Green:2026yev} for recent studies searching for cosmological collider signals in cosmological data), for many aforementioned models their resulting bispectra have not been computed in the full physical kinematic space. This is insufficient for performing an accurate search using current cosmological data, where the non-squeezed region constitute a large fraction of the available statistics. 
Moreover, naively extrapolating the oscillatory signature to the full kinematic space is misleading, since the off-shell and non-oscillatory background piece is known to dominate more strongly in the non-squeezed region. 
This also means many theoretically well-motivated processes have not been searched for in cosmological data, including DE, TE, and SCP.

In an accompanying work~\cite{Kumar:2026ogn}, we have started to fill this gap between the theoretical development of cosmological collider and the practicality of observational searches. 
In this longer work, we present efficient computation of bispectrum shape for the multiple exchange and the SCP scenario in the entire kinematic space and across a range of model parameters. 
Together with~\cite{Kumar:2026ogn}, this work provides a unified treatment of SE, DE, and TE processes, and the first computation of full-kinematic bispectrum for the TE and the SCP processes. 
The observational motivation for the SCP model is immediately clear from its ability to counter the exponential suppression, as discussed before. 
The TE diagram is also observationally more motivated than its SE and DE counterparts for two reasons.

First, for the TE topology, looking at the squeezed limit of the inflaton bispectrum (e.g., $k_3 \ll k_1, k_2$) is equivalent to the limit where the momentum of the exchanged scalar is small, since every external inflaton legs has a two-point mixing with the heavy scalar. 
This squeezed limit corresponds to the on-shell propagation of the heavy scalar and gives the distinctive oscillatory dependence in the bispectrum (eq.~\eqref{eq:ccp_schematic}).
On the other hand, for SE and DE topologies, the squeezed limit of the inflaton bispectrum is always accompanied by {\it non-oscillatory} contributions from the permutation diagrams where $k_3$ is assigned to the inflaton leg that does not have a two-point mixing with $\sigma$. 
In this case, the momentum of $\sigma$ is not small, compared to $H$, and these permutations do not correspond to the on-shell propagation of $\sigma$.
Thus, for both SE and DE, the oscillatory contributions are less distinctive since the full bispectrum is a sum of oscillatory and non-oscillatory pieces in the squeezed limit.
However, for TE, there are no non-oscillatory contributions in the squeezed limit.
Thus, the oscillatory feature in the bispectrum is the most prominent in a TE topology than either SE or DE. 

Second, in TE, the bispectrum is enhanced by extra insertion of the homogeneous inflaton rolling $\dot{\phi}_0$ and therefore results in a parametrically larger NG than SE and DE.
We will demonstrate this by translating the NG constraints to model parameters. 
In fact, we will show that the {\it Planck} data currently does not constrain either SE or DE processes within the perturbative limit, while it does place meaningful constraints on TE processes.\footnote{Even in the non-perturbative regime of the quadratic mixing, the TE process is expected to give a parametrically larger signal and be most meaningfully probed by the current data~\cite{Kumar:2026ogn}.}

With the full bispectrum shapes computed, we search for SE, DE, TE, and SCP signatures in the {\it Planck} data using the {\tt CMB-BEST} pipeline~\cite{Sohn:2023fte}. 
As in the accompanying work~\cite{Kumar:2026ogn}, the SCP mechanism allows us to do the first search of unsuppressed oscillatory bispectrum mediated by particles with $M\gg H$, since the previous searches in the literature used templates where the oscillatory signatures gets exponentially suppressed for $M\gg H$ and the bispectrum effectively reduces to the equilateral shape.
% In the $M\gg H$ regime, the SCP model in particular presents the first search of the oscillatory signature, since previous work only considered the SE~\cite{?}, and to a limited extent DE~\cite{Philcox:2025bbo}, process where the oscillatory signal is Boltzmann-suppressed in the large mass regime and the bispectrum is effectively of the equilateral shape. 
In the $M\sim H$ regime, we are providing a broader search of DE processes, and first searches for TE and SCP processes.
%As discussed previously, all these three processes feature much more prominent oscillatory bispectra compared to the SE process previously studied in the literature.

{\bf Notation:} In this work, we work with mostly plus metric convention $(- + + +)$. We have used $H=1$ units in equations, except for the discussions where $H$ is kept explicitly to clarify various parametric dependence.

The rest of this work is organized as follows. In Sec.~\ref{sec:treefNL}, we start by explaining the computational methodology to obtain the full shape of the bispectrum.
We then present the model Lagrangians we consider in this work and discuss salient features in the bispectrum shapes in each model. 
In Sec.~\ref{sec:Planck} we present the result from searching for each shape in \textit{Planck} data using the \texttt{CMB-BEST} code. 
% We demonstrate that there is a hint of data preference for unsuppressed oscillation in the shape of non-Gaussianities.
We conclude and discuss future directions in Sec.~\ref{sec:discussion}.

\section{Tree-level Bispectrum from Massive Scalars}\label{sec:treefNL}

In this section we first review the computational method used in our work. We then discuss the models of interest in more detail, specify their Lagrangian, and present the bispectrum shape from each model in the full kinematic space. We also provide the benchmark effective field theory (EFT) interpretation for each model and discuss constraints on the model parameter space from perturbativity and power-spectrum corrections.

\subsection{Computation Methodology}\label{sec:method}

In this work, we utilize two different computational methods to evaluate the bispectrum from SE, DE, TE, and SCP. The first is the Cosmological Bootstrap method~\cite{Arkani-Hamed:2018kmz, Baumann:2019oyu}, where a differential equation, that the correlator satisfies, is solved analytically. The second is the Coupled Model Function (CMF) approach~\cite{An:2017hlx}, where the effective propagator from the inflaton-heavy scalar two-point mixing is numerically computed.
This reduces the exchange diagrams to contact diagrams, which can then be easily numerically integrated over Euclidean de Sitter (dS) time for each external momentum configuration. In the rest of the section we provide a review of both methodologies, but first we will briefly explain the choice of computational method for the models considered in this paper. 

{\bf Summary of methods adopted in this work:} The bootstrap method, in principle, produces an analytic solution across all kinematic space. The SE diagram has been studied most extensively in the literature using bootstrap, and an analytical closed form solution was derived in~\cite{Qin:2023ejc}. For DE, the bootstrap method becomes substantially more complex, and a series expansion in small $k_3/k_1$ has been found in~\cite{Xianyu:2023ytd, Aoki:2024uyi}. The bootstrap method, for either model, is however numerically unstable near the folded limit $k_1+k_2\approx k_3$ because of cancellation of superfluous divergences. 
For SE, this does not prevent the bootstrap method from practical use since the numerical instability occurs outside the range of momenta configurations probed by the current data (see Fig.~\ref{fig:single_folded}).
However, for DE the numerical instability is severe (see Fig.~\ref{fig:double_folded}). 
For TE, the bootstrap solution has not yet been found. For these reasons, we compute the SE, DE, and TE diagrams in a unified framework using CMF. We have verified that CMF and bootstrap produce the same result for SE, and in numerically stable regions, for DE (see Appendix~\ref{app_numerical}). Even for SE, we also find that evaluating the analytic formula from the bootstrap method on an external momentum triangle is not meaningfully faster than numerically evaluating the diagram using the CMF method. 

The SCP scenario, unlike the multiple exchange scenarios, has three degrees of freedom since the heavy scalar is necessarily a complex field to carry an $U(1)$ current. The extra degree of freedom makes the CMF method more involved. Therefore, in this work, to evaluate the SCP scenario, we use the bootstrap result from~\cite{Qin:2023ejc}, accounting for the appropriate inflaton-scalar coupling.

\subsubsection{Cosmological Bootstrap}\label{sec:bootstrap}
The central idea behind the Bootstrap approach is to derive a set of differential equations that the cosmological correlators satisfy. These equations can then be solved, subject to boundary conditions, to obtain the general kinematic form of the correlators.

To illustrate the basic idea, consider the SE diagram with some interaction between the inflaton and the heavy scalar. Recall that a massive free scalar field $\Phi(\eta,\vec{k})$ in dS satisfies the Klein-Gordon equation, 
\es{}{
\eta^2\partial_\eta^2\Phi - 2\eta\partial_\eta \Phi + (k\eta)^2\Phi + M^2\Phi \equiv {\cal D}(\eta,k)\Phi = 0.
}
By solving this equation, the mode functions $f$ and $\bar{f}$ for the massive field in dS can be derived.
Canonical quantization follows by expressing the field operator in terms of the creation and annihilation operators, and the mode functions,
\es{}{
\Phi(\eta,\vec{k}) = a_{\vec{k}}f_k(\eta) + a_{-\vec{k}}^\dagger \bar{f}_k(\eta).
}
To obtain the mode functions for the inflaton fluctuations $\varphi$, which we can approximate as a massless field, the above equation can be solved with $M\approx 0$.
We also recall the master formula for the `in-in' expectation value of an operator ${\cal Q}$
\es{eq:in_in}{
\langle {\cal Q}\rangle = \langle 0 | {\cal U}^\dagger {\cal Q}_I(t_{\rm f}) {\cal U} |0\rangle,
}
where ${\cal U} = T\exp(-\ii\int_{-\infty(1-\ii\epsilon)}^{t_{\rm f}} \D t H_I^{\rm int}(t))$ is the time evolution operator obtained from the interacting part of the interaction picture Hamiltonian $H_I^{\rm int}$, and $t_{\rm f}$ is a time towards the end of inflation.
With the above mode functions and the `in-in' formula in Eq.~\eqref{eq:in_in}, the diagram where $k_3$ flows through the exchanged $\sigma$ (denoted by the subscript $\sigma_{\vec{k}_3}$) can be written as,
\es{eq:3pt_single_schematic}{
&\langle \varphi(\vec{k}_1)\varphi(\vec{k}_2)\varphi(\vec{k}_3)\rangle_{\sigma_{\vec{k}_3}} \\
&= \sum_{a,b=\pm} (-ab) \int_{-\infty}^0 {\D \eta_1 \over (-\eta_1)^4} \int_{-\infty}^0 {\D \eta_2 \over (-\eta_2)^4}\\
&\times [{\cal O}_1 G_a(\eta_1, k_1)]\cdot [{\cal O}_2 G_a(\eta_1, k_2)]\cdot [{\cal O}_3 G_b(\eta_2, k_3)] \\& \times  D_{ab}(\eta_1, \eta_2, k_3)V_1(\eta_1)V_2(\eta_2) + {\rm perms.},
}
where the operators ${\cal O}$ depend on the various $\varphi-\sigma$ interaction terms in the Lagrangian, and $V_i$ contain various model-dependent vertex factors.
We give explicit examples of ${\cal O}$ and $V_i$ below.
The indices $a$ and $b$ denote whether the vertices come from time-ordering ($+$) or anti-time ordering ($-$), and the overall minus sign is the $(-\ii)^2$ factor from the time evolution operator.
The inflaton bulk-to-boundary propagators are given by,
\es{}{
G_a(\eta, k) = {1\over 2k^3} (1-\ii a k \eta)e^{\ii ak\eta},
}
which satisfy the relation $\partial_\eta G_a(\eta, k) = \eta e^{\ii ak\eta}/(2k)$, often useful in explicit evaluation.
The bulk massive propagators are denoted by $D_{ab}$,
\es{}{
D_{++}(\eta_1, \eta_2, k) &= f_k(\eta_1) \bar{f}_k(\eta_2) \theta(\eta_1-\eta_2) \\&+ f_k(\eta_2) \bar{f}_k(\eta_1) \theta(\eta_2-\eta_1),\\
D_{--}(\eta_1, \eta_2, k) &= f_k(\eta_1) \bar{f}_k(\eta_2) \theta(\eta_2-\eta_1) \\&+ f_k(\eta_2) \bar{f}_k(\eta_1) \theta(\eta_1-\eta_2),\\
D_{+-}(\eta_1, \eta_2, k) &= f_k(\eta_2) \bar{f}_k(\eta_1),\\
D_{-+}(\eta_1, \eta_2, k) &= f_k(\eta_1) \bar{f}_k(\eta_2).
}
 The evaluation of $\langle \varphi(\vec{k}_1)\varphi(\vec{k}_2)\varphi(\vec{k}_3)\rangle_{\sigma_{\vec{k}_3}}$ can be done by first computing a `seed integral'~\cite{Qin:2023ejc},
\es{seed}{
&{\cal I}_{ab}^{p_1p_2}(u_1,u_2) \equiv (-ab)k_s^{5+p_{12}}\int_{-\infty}^0 \D\eta_1\int_{-\infty}^0\D\eta_2(-\eta_1)^{p_1}\\ &\times(-\eta_2)^{p_2}e^{\ii ak_{12}\eta_1+\ii bk_{34}\eta_2}D_{ab}(\eta_1,\eta_2, k_s),
}
where $u_1= 2k_s/(k_{12}+k_s)$, $u_2= 2k_s/(k_{34}+k_s)$ and $p_{12}=p_1+p_2$, and then taking into account the actions of the operators ${\cal O}_{1,2,3}$.
In the present example, $k_s=k_3$, $k_4=0$, resulting in $u_2=1$.
Now, given the Klein-Gordon equation, it follows,
\es{}{
{\cal D}(\eta_1,k) D_{\pm \mp}(\eta_1, \eta_2, k) &= 0,\\
{\cal D}(\eta_1,k) D_{\pm \pm}(\eta_1, \eta_2, k) &= \mp \ii\eta_1^2 \eta_2^2 \delta(\eta_1-\eta_2).
}
Using these equations, one can derive the differential equations satisfied by ${\cal I}_{ab}^{p_1p_2}(u_1,u_2)$.
Upon solving those equations, subject to certain boundary conditions, one can obtain an expression for $\langle \varphi(\vec{k}_1)\varphi(\vec{k}_2)\varphi(\vec{k}_3)\rangle_{\sigma_{\vec{k}_3}}$. The resulting expression for ${\cal I}_{ab}^{p_1p_2}(u_1,1)$ for arbitrary values of $p_1$ and $p_2$ can be found in~\cite{Qin:2023ejc}.

So far we have computed the contribution to the three-point inflaton correlator where $\vec{k}_3$ flows through the exchanged massive particle $\sigma$. The full three-point inflaton correlator is a sum over all possible momentum assignment choices. Assuming $\langle \varphi(\vec{k}_1)\varphi(\vec{k}_2)\varphi(\vec{k}_3)\rangle_{\sigma_{\vec{k}_3}}$ is symmetric upon $\vec{k}_1 \leftrightarrow \vec{k}_2$, as will be the case for the examples below, we only need to sum over cyclic permutations of the three external momenta,
\es{eq:SE_full}{
&\langle \varphi(\vec{k}_1)\varphi(\vec{k}_2)\varphi(\vec{k}_3)\rangle = \langle \varphi(\vec{k}_1)\varphi(\vec{k}_2)\varphi(\vec{k}_3)\rangle_{\sigma_{\vec{k}_3}}\\ &+ \langle \varphi(\vec{k}_2)\varphi(\vec{k}_3)\varphi(\vec{k}_1)\rangle_{\sigma_{\vec{k}_1}}+\langle \varphi(\vec{k}_3)\varphi(\vec{k}_1)\varphi(\vec{k}_2)\rangle_{\sigma_{\vec{k}_2}}.
}
The sum over momentum permutation is why the squeezed limit of the bispectrum, $\lim_{k_3/k_1\rightarrow 0} \langle \varphi(\vec{k}_1)\varphi(\vec{k}_2)\varphi(\vec{k}_3)\rangle$, necessarily contains not only the observationally interesting oscillatory signature from the first term in Eq.~\eqref{eq:SE_full}, where momentum of $\sigma$ is small, but also the smooth ``background'' from the last two terms where it is not. We will compute these background contributions explicitly when we discuss the actual models and their bispectrum shapes in Sec.~\ref{sec:shape}.

\subsubsection{Coupled Mode Function}\label{sec:CMF}
Besides the bootstrap method which provides analytical results for certain diagrams, we will introduce the Coupled Mode Function (CMF) method as an efficient numerical approach for evaluating cosmological correlators. This method deals with the quadratic mixing of fields non-perturbatively, resulting in a set of coupled mode functions. With this set of functions, the evaluation of multiple exchange diagrams in Fig. \ref{fig:topo} turns into the numerical integration of 3-point contact diagrams. Because of its non-perturbative nature, it can be applied not only to the strongly mixing regime \cite{An:2017hlx}, but also to time-dependent backgrounds \cite{Reece:2022soh}.\par
The key idea of this method is to treat the quadratic and cubic interactions of fields in different steps. We suppose that our theory is defined by the Lagrangian of the form ${\cal L}={\cal L}_2+{\cal L}_{\rm int}$, in which ${\cal L}_2$ consists of quadratic terms including the mixing term, and ${\cal L}_{\rm int}$ is the interaction term consisting of cubic and higher-order couplings. For concreteness, consider
\es{}{
\mathcal{L}_2=-\frac{1}{2}a^4(\partial\varphi)^2-\frac{1}{2}a^4(\partial\sigma)^2-\frac{1}{2}a^4 m_{\sigma}^2\sigma^2+\lambda_2 a^3\varphi'\sigma.
}
The equations of motion in dS spacetime, in which $a=-1/\eta$, are given by
\begin{equation}\label{coupled_eom}
    \begin{aligned}
        &\varphi_{\bf k}''-\frac{2}{\eta}\varphi_{\bf k}'+k^2\varphi_{\bf k}=\lambda_2\left(\frac{1}{\eta}\sigma_{\bf k}'-\frac{3}{\eta^2}\sigma_{\bf k}\right),\\
        &\sigma_{\bf k}''-\frac{2}{\eta}\sigma_{\bf k}'+\left(k^2+\frac{m_{\sigma}^2}{\eta^2}\right)\sigma_{\bf k}=-\frac{\lambda_2}{\eta}\varphi_{\bf k}',
    \end{aligned}
\end{equation}
in which $\varphi_{\bf k}$ and $\sigma_{\bf k}$ are Fourier modes of fields in 3-momentum space. This is a set of coupled equations that we solve numerically, subject to appropriate initial conditions. To do this, we first note that this set of equations has two sets of independent solutions that we may denote as $\Phi^{(1)}_{k}=\left(\varphi_{k}^{(1)},\sigma_{k}^{(1)}\right)$ and $\Phi^{(2)}_{k}=\left(\varphi_{\bf k}^{(2)},\sigma_{k}^{(2)}\right)$, and the full solution can be written as
\begin{equation}
    \Phi_{\bf k}(\eta)=\begin{pmatrix}\varphi_{\bf k}(\eta)\\\sigma_{\bf k}(\eta)\end{pmatrix}=a^{(1)}_{\bf k}\Phi^{(1)}_k(\eta)+a^{(2)}_{\bf k}\Phi^{(2)}_k(\eta)+{\rm h.c.}.
\end{equation}
Now we need to apply canonical quantization to this solution. Given the field variables being $\varphi$ and $\sigma$, the corresponding canonical momenta are
\begin{equation}
    \begin{aligned}
        \pi_\varphi&=a^2\varphi'+\lambda_2 a^3\sigma,\\
        \pi_\sigma&=a^2\sigma'.
    \end{aligned}
\end{equation}
The canonical commutation relations require that $[\varphi,\pi_\varphi]=[\sigma,\pi_\sigma]=\ii$ and all other commutators be zero. Due to the couplings, the commutators of $a^{(1,2)}_{\bf k}$ and $a^{(1,2)\dagger}_{\bf k}$ derived from these relations will generally be complicated. However, in the early time limit $\eta\to-\infty$, we note that the equation reduces to that of two decoupled massless scalars and $\pi_\varphi$ reduces to $a^2\varphi'$. Therefore, in this limit we require that the commutators satisfy
\begin{equation}
    \left[a^{(i)}_{\bf k},a^{(j)\dagger}_{\bf q}\right]
    =(2\pi)^3\delta^{(3)}({\bf k}-{\bf q})\delta_{ij}.
\end{equation}
The canonical commutation relations $[\varphi,\pi_\varphi]=\ii$ and $[\varphi,\sigma]=0$ lead to the following conditions:
\begin{equation}\label{wrongski}
    \begin{aligned}
        \sum_{i=1,2}\left[\varphi^{(i)}_{k}\left(\partial_\eta\varphi^{(i)*}_{k}\right)-\varphi^{(i)*}_{k}\left(\partial_\eta\varphi^{(i)}_{k}\right)\right]&=\frac{\ii}{a^2},\\
        \sum_{i=1,2}\left(\varphi^{(i)}_{k}\sigma^{(i)*}_{k}-\varphi^{(i)*}_{k}\sigma^{(i)}_{k}\right)&=0.
    \end{aligned}
\end{equation}
These conditions determine the correct normalization of the mode functions $\Phi^{(1)}_{k}$ and $\Phi^{(2)}_{k}$.\footnote{We remark here that while the two condition above is sufficient to determine the prefactors of $\Phi^{(1)}_{k}$ and $\Phi^{(2)}_{k}$, the other commutation relations $[\varphi,\pi_{\sigma}]=[\pi_{\varphi},\sigma]=[\pi_\varphi,\pi_\sigma]=0$ leave nontrivial constrains on the functional form of them. We have checked that our choice of $\Phi^{(1)}_{k}$ and $\Phi^{(2)}_{k}$ respects these constraints.}\par
\begin{figure}[htbp]
    \centering
\includegraphics[width=1\linewidth]{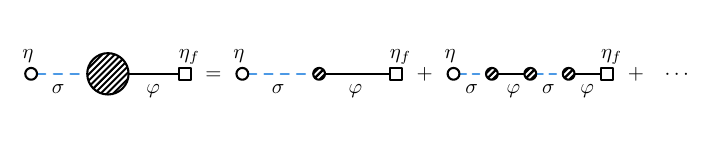}
    \caption{An illustration of a CMF bulk-to-boundary propagator that connects a $\sigma$ field at bulk time $\eta$ and a $\phi$ field at boundary time $\eta_f$. It is a resummation of a series of two-point correlators with multiple insertions of quadratic vertices. In these expression, the time integrals and Schwinger-Keldysh summations on shaded vertices on the RHS has already been completed.}
    \label{fig:resummed}
\end{figure}

To get the explicit form of $\Phi^{(1)}_{k}$ and $\Phi^{(2)}_{k}$ in the early time limit as our initial conditions, we neglect the terms proportional to $1/\eta^2$ in (\ref{coupled_eom}) and get
\begin{equation}
    \begin{aligned}
        \varphi_{\bf k}''-\frac{2}{\eta}\varphi_{\bf k}'+k^2\varphi_{\bf k}&=\frac{\lambda_2}{\eta}\sigma_{\bf k}',\\
        \sigma_{\bf k}''-\frac{2}{\eta}\sigma_{\bf k}'+k^2\sigma_{\bf k}&=-\frac{\lambda_2}{\eta}\varphi_{\bf k}'.
    \end{aligned}
\end{equation}
In the limit of $\eta\to-\infty$, the $k^2\varphi_{\bf k}$ and $k^2\sigma_{\bf k}$ terms dominate the oscillations, which motivates us to take the following ansatz
\begin{equation}\label{ansatz}
    \begin{aligned}
        \Phi_{k}(\eta)=\begin{pmatrix}A_k(\eta)\\B_k(\eta)\end{pmatrix}e^{-\ii k\eta},
    \end{aligned}
\end{equation}
in which we have $A'_k/A_k, B'_k/B_k\ll1$. Then to the leading order, the equations become
\begin{equation}
    \begin{aligned}
        2A'_k-\frac{2}{\eta}A_k-\frac{\lambda_2}{\eta}B_k=&0,\\
        2B'_k-\frac{2}{\eta}B_k+\frac{\lambda_2}{\eta}A_k=&0.
    \end{aligned}
\end{equation}
This set of equation can be solved to get
\begin{equation}
    \begin{aligned}
        A^{(1)}_k=&c_1(-\eta)^{1+\ii\frac{\lambda_2}{2}},\quad B^{(1)}_k=\ii A^{(1)}_k,\\
        A^{(2)}_k=&c_2(-\eta)^{1-\ii\frac{\lambda_2}{2}},\quad B^{(2)}_k=-\ii A^{(2)}_k.
    \end{aligned}
\end{equation}
Inserting the above result back to (\ref{wrongski}), we get the value of normalization coefficients $c_1$ and $c_2$, and the canonically normalized solution reads
\begin{equation}\label{init}
    \begin{aligned}
        \varphi^{(1)}_k=&\frac{1}{2k^{3/2}}(-k\eta)^{1+\ii\frac{\lambda_2}{2}}e^{-\ii k\eta},\quad\sigma^{(1)}_k=\ii\varphi^{(1)}_k,\\
        \varphi^{(2)}_k=&\frac{1}{2k^{3/2}}(-k\eta)^{1-\ii\frac{\lambda_2}{2}}e^{-\ii k\eta},\quad\sigma^{(2)}_k=-\ii\varphi^{(2)}_k.\\
    \end{aligned}
\end{equation}
This gives the initial condition required to solve the full equation (\ref{coupled_eom}). To proceed, we still use the ansatz (\ref{ansatz}) and solve $A_k(\eta)$ and $B_k(\eta)$ numerically. More specifically, by defining $z\equiv-\ii k\eta$, we introduce the Wick-rotated equations of motion:
\begin{equation}
    \begin{aligned}
        A_k''(z)+\frac{2(z-1)}{z}A'_k(z)-\frac{2}{z}A_k(z)-\frac{\lambda_2}{z}B'_k(z)\\ -\frac{\lambda_2(z-3)}{z^2}B_k(z)=0,\\
        B''_k(z)+\frac{2(z-1)}{z}B'_k(z)+\frac{m^2-2z}{z^2}B_k(z)\\+\frac{\lambda_2}{z}\left(A'_k(z)+A_k(z)\right)&=0.
    \end{aligned}
\end{equation}
These are the equations we use to obtain the numerical solution, with the initial condition Wick-rotated accordingly (See \cite{Reece:2022soh} for more explanations). From the solution of coupled mode functions, we can construct the two-point correlator $\langle\sigma_{\bf k}(\eta)\varphi_{\bf k}(\eta_f)\rangle'$ as follows: 
\begin{equation}
    \begin{aligned}
        \langle\sigma_{\bf k}(\eta)\varphi_{\bf k}(\eta_f)\rangle'&=\left\langle\left[a^{(1)}_{\bf k}\sigma^{(1)}_k(\eta)+a^{(2)}_{\bf k}\sigma^{(2)}_k(\eta)+{\rm h.c.}\right]\right.\\ & \times\left.\left[a^{(1)}_{\bf k}\varphi^{(1)}_k(\eta_f)+a^{(2)}_{\bf k}\varphi^{(2)}_k(\eta_f)+{\rm h.c.}\right]\right\rangle\\
        &=\sigma^{(1)}_k(\eta)\varphi^{(1)*}_k(\eta_f)+\sigma^{(2)}_k(\eta)\varphi^{(2)*}_k(\eta_f),
    \end{aligned}
\end{equation}
which is effectively a tree-level resummed correlator, illustrated in Fig.~\ref{fig:resummed}. Taking $\eta_f$ to be on the late time slice during inflation, it becomes the $(-)$-type bulk-to-boundary propagator in the in-in formalism. We will take $\eta_f=0$ for clarity in the following context.

So far we have encoded the effect of quadratic mixing in the form of $\langle\sigma_{\bf k}(\eta)\varphi_{\bf k}(0)\rangle'$. The three-point correlators, given by cubic terms in ${\cal L}_{\rm int}$, can now be calculated simply by numerical integration. For example, the SE diagram provided by ${\cal L}_{\rm int}\supset\frac{1}{2}\lambda_{3}^{(1)} a^2\varphi'^2\sigma$, where $k_3$ is the momentum of the exchanged particle, is given as\footnote{Note that in numerical calculation the limits of integration are taken as finite values $\eta_i$ and $\eta_f$, which should taken to be sufficiently large and small (in magnitude), respectively, for the desired level of numerical accuracy.}
\es{cmf_single}{
    &\langle\varphi_{\bf k_1}\varphi_{\bf k_2}\varphi_{\bf k_3}\rangle'_{\sigma_{\bf k_3}}=2\lambda_{3}^{(1)}\times\\& \Re\left[\int_{-\infty}^0\frac{\dd\eta}{\eta^2}G'_-(\eta,k_1)G'_-(\eta,k_2)\langle\sigma_{{\bf k}_3}(\eta)\varphi_{{\bf k}_3}(0)\rangle'\right],
}
where $G'_-(\eta,k) = \partial_\eta G_-(\eta,k)$.
Note that although the quadratic coupling $\lambda_2$ does not appear in the above expression at first glance, the coupled mode function $\langle\sigma_{\bf k}(\eta)\varphi_{\bf k}(0)\rangle'$ has implicit dependence on it. As discussed at the end of Sec~\ref{sec:bootstrap}, the full three-point inflaton correlator is then the cyclic permutation over external momentum assignments:
\es{}{
&\langle \varphi(\vec{k}_1)\varphi(\vec{k}_2)\varphi(\vec{k}_3)\rangle_{\rm SE} = \langle \varphi(\vec{k}_1)\varphi(\vec{k}_2)\varphi(\vec{k}_3)\rangle_{\sigma_{\vec{k}_3}}\\&+ \langle \varphi(\vec{k}_2)\varphi(\vec{k}_3)\varphi(\vec{k}_1)\rangle_{\sigma_{\vec{k}_1}}+\langle \varphi(\vec{k}_3)\varphi(\vec{k}_1)\varphi(\vec{k}_2)\rangle_{\sigma_{\vec{k}_2}}.
}

Similarly, the DE and TE diagrams provided by ${\cal L}_{\rm int}\supset\frac{1}{2}\lambda_{3}^{(2)} a^3\varphi'\sigma^2+\frac{1}{6}\lambda_{3}^{(3)} a^4\sigma^3$ are given respectively as
\es{cmf_double}{
&\langle\varphi_{\bf k_1}\varphi_{\bf k_2}\varphi_{\bf k_3}\rangle'_{\sigma_{\bf k_1}\sigma_{\bf k_3}}=2\lambda_{3}^{(2)}\times \\
    &\Re\left[\int_{-\infty}^0\frac{\dd\eta}{\eta^3} G'_-(\eta,k_2)\langle\sigma_{{\bf k}_1}(\eta)\varphi_{{\bf k}_1}(0)\rangle'\langle\sigma_{{\bf k}_3}(\eta)\varphi_{{\bf k}_3}(0)\rangle'\right],
}
and
\es{cmf_triple}{
\langle\varphi_{\bf k_1}\varphi_{\bf k_2}\varphi_{\bf k_3}&\rangle'_{\sigma_{\bf k_1}\sigma_{\bf k_2}\sigma_{\bf k_3}}=2\lambda_{3}^{(3)}\times \\&\Re\left[\int_{-\infty}^0\frac{\dd\eta}{\eta^4}\prod_{j=1}^3\langle\sigma_{{\bf k}_j}(\eta)\varphi_{{\bf k}_j}(0)\rangle'\right].}
For the DE diagram, again we need to sum over all possible external momentum assignment choices to get the full bispectrum,
\es{}{
&\langle \varphi(\vec{k}_1)\varphi(\vec{k}_2)\varphi(\vec{k}_3)\rangle_{\rm DE} = \langle\varphi_{\bf k_1}\varphi_{\bf k_2}\varphi_{\bf k_3}\rangle'_{\sigma_{\bf k_1}\sigma_{\bf k_3}} \\&+\langle\varphi_{\bf k_2}\varphi_{\bf k_3}\varphi_{\bf k_1}\rangle'_{\sigma_{\bf k_2}\sigma_{\bf k_1}} +\langle\varphi_{\bf k_3}\varphi_{\bf k_1}\varphi_{\bf k_2}\rangle'_{\sigma_{\bf k_3}\sigma_{\bf k_2}}. 
}
In the squeezed limit $k_3/k_1\ll 1$, the first and the third term will result in oscillatory features, while the second term will lead to a smooth ``background'' contribution.

For the TE diagram however, there is no need to sum over momentum assignments, since the single diagram we computed is already fully symmetric with respect to $(\vec{k}_1, \vec{k_2}, \vec{k}_3)$:
\begin{equation}
    \langle \varphi(\vec{k}_1)\varphi(\vec{k}_2)\varphi(\vec{k}_3)\rangle_{\rm TE} = \langle\varphi_{\bf k_1}\varphi_{\bf k_2}\varphi_{\bf k_3}\rangle'_{\sigma_{\bf k_1}\sigma_{\bf k_2}\sigma_{\bf k_3}},
\end{equation}
and squeezing the momentum triangle will only result in oscillatory features, unaffected by smooth contributions.

\iffalse
In each time of numerical integration, we can only calculate the value of a diagram with certain \textit{fixed} configuration $\left(k_1,k_2,k_3\right)$ of momenta, which means that we can only obtain the value of a diagram on a discrete grid in configuration space and a continuous numerical function is obtained by smooth interpolation.\par

In summary, the CMF method can provide numerical templates in the form of interpolating functions for various theories with both quadratic and cubic couplings. We make a few remarks here. 1) The CMF method can be applied naturally to the regime of strongly two-point mixing, as originally emphasized by \cite{An:2017hlx}, and we will use it to derive templates of scalar exchanges in this regime. 2) For scale-invariant theories, the scaling law of diagrams helps to reduce the required number of independent integrations necessary for interpolations, greatly enhancing the efficiency of calculation, while for the case of breaking scale symmetry the procedure can be much more time consuming. 3) This method can be applied to the case with time-dependent couplings with minor changes in intermediate steps, as we will illustrate in the case of primordial features. 
\fi

\subsection{Models and Shapes}\label{sec:shape}

In this subsection we specify the Lagrangian for each model, discuss benchmark constraints on the model parameters from perturbativity, and present important characteristics of the bispectrum shape obtained using the computational method reviewed above. Given a three-point correlator $\langle\varphi_{\bf k_1}\varphi_{\bf k_2}\varphi_{\bf k_3}\rangle'$, we define the dimensionless shape function by
\begin{equation}
    S(k_1,k_2,k_3)=(k_1 k_2 k_3)^2\langle\varphi_{\bf k_1}\varphi_{\bf k_2}\varphi_{\bf k_3}\rangle'.\label{eq:shapedef}
\end{equation}
Unless explicitly specified, $\langle\varphi_{\bf k_1}\varphi_{\bf k_2}\varphi_{\bf k_3}\rangle'$ is the complete three-point correlator after summing over all external momentum permutations. Furthermore, the shape function is manifestly scale-invariant for all models we consider in this work, since the interactions respect the inflaton shift symmetry $\varphi\rightarrow\varphi+{\rm constant}$. For both the CMF and the bootstrap method, we obtain a continuous function in the $(k_1, k_2, k_3)$ space by calculating $S(k_1, k_2, k_3)$ on a lattice in the $(\log k_1, \log k_2, \log k_3)$ space and interpolating. Since the frequency of the oscillatory signal is dictated by the mass parameter $\tilde{\nu} \equiv \sqrt{M^2/H^2-9/4}$, the spacing of the lattice should be much smaller than $2\pi/\tilde{\nu}$ to capture the oscillations.

For all models studied in this paper, we impose the constraint that the correction to the inflaton power spectrum due to the interaction between the inflaton and the new heavy scalar, $\Delta P_{\zeta}$, is smaller than the power spectrum from the inflaton potential, $P_{\zeta}$, which we take to match the observed power spectrum amplitude. This is not a fundamental constraint, since observationally we cannot distinguish $\Delta P_{\zeta}$ and $P_{\zeta}$ due to their scale invariance. However, considering the regime $\Delta P_{\zeta} > P_{\zeta}$ requires resumming the two-point mixing between the inflaton and the new scalar, and we leave the detailed treatment of this regime for future work.

\subsubsection{Single Exchange}\label{sec:SiEx}
Firstly we consider a model for the SE diagram, where the non-Gaussianity is sourced by the scale-invariant couplings
\begin{equation}\label{SE_model}
    {\cal L}_{\rm int}\supset\tilde{\lambda}_2\Lambda a^3\varphi'\sigma+\frac{\tilde{\lambda}_3^{(1)}}{2\Lambda}a^2\varphi'^2\sigma,
\end{equation}
in which we have introduced a cutoff scale $\Lambda$ to make the coupling constants $\tilde{\lambda}$ dimensionless. 

\begin{figure}[htbp]
    \centering
    \includegraphics[width=0.9\linewidth]{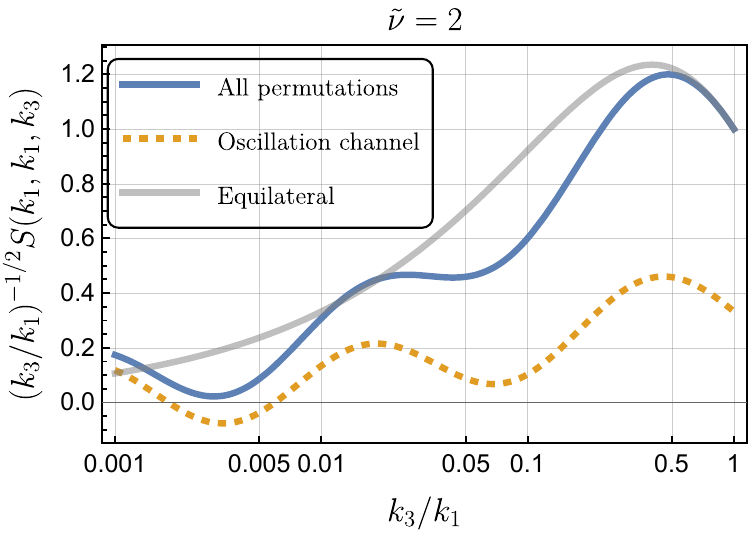}
    \hspace{0.018\linewidth}
    \includegraphics[width=0.9\linewidth]{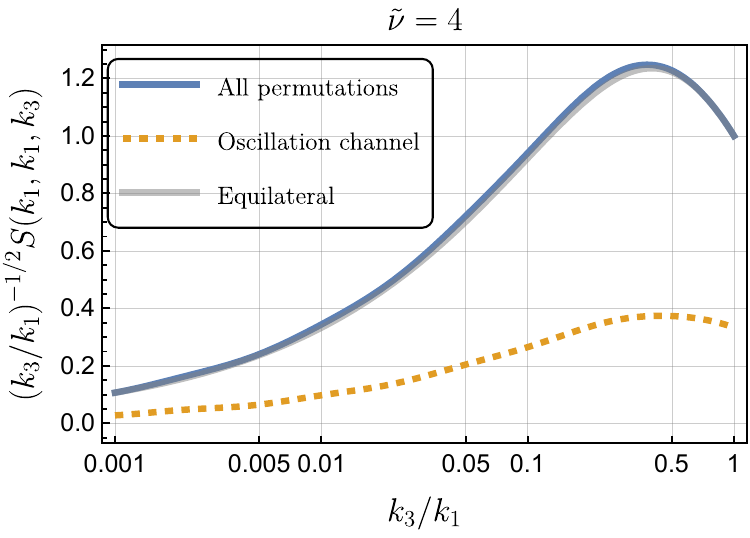}
    \caption{The shape function of the single exchange model, with the diagram corresponding to the oscillation channel (dashed yellow) and sum of all external momentum permutations (solid blue), normalized to unity at equilateral configuration. We have plotted the standard equilateral shape (solid gray) for comparison. We can see that the permutations give a large, non-oscillatory contribution to the total shape at non-squeezed region. At large mass, the signal component is obscured and the total shape becomes indistinguishable with the equilateral shape.}
    \label{fig:single_all}
\end{figure}

In Fig.~\ref{fig:single_all}, we show the shape functions in isosceles configuration $S(k,k,k_3)$ for various masses calculated with the CMF method. We show the result (dashed yellow) from the single diagram in which the momentum of the massive propagator is $\mathbf{k}_3$ and which contributes to oscillatory signals in the squeezed limit $k_3/k\to0$. We call this the `oscillation channel'. We also show the sum of all diagrams accounting for permutations (solid blue). For comparison, we also plot the standard equilateral shape defined by
\begin{equation}\label{equil}
    S_{\rm{equil}}(k_1,k_2,k_3)=\frac{27k_1k_2k_3}{(k_1+k_2+k_3)^3}.
\end{equation}
This type of shape is produced by the effective coupling term $\varphi'^3$, which is obtained by integrating out the heavy particle in the multiple exchange models. 

From the oscillation channel curve in Fig.~\ref{fig:single_all}, we see that the frequency of the oscillations is proportional to $\tilde{\nu}$. 
As $\tilde{\nu}$ becomes large, the off-shell background shape dominates over the oscillations due to the Boltzmann suppression of on-shell particle productions. 
From the sum of all permutations, we see that permutations other than the oscillation channel contribute only to a non-oscillatory background. As the mass increases, the signal is strongly suppressed and the full shape becomes indistinguishable from the equilateral shape.\par

{\bf Benchmark EFT Interpretation:}
The interactions in~\eqref{SE_model} can originate from a Lorentz-invariant dimension-5 operator,
\es{eq:SE_eft}{
\sqrt{-g}{(\partial \phi)^2\sigma \over \eftcut}\supset {1 \over \eta^4\eftcut } (2\eta\dot{\phi}_0 \varphi'\sigma -\eta^2 \varphi'^2 \sigma) +\cdots.
}
This gives $\tilde{\lambda}_2\Lambda = -2\dot{\phi}_0/\eftcut$ and $\tilde{\lambda}_3^{(1)}/(2\Lambda) = -1/\eftcut$.
Thus the product that controls the shape function is given by,
\es{}{
\tilde{\lambda}_2 \tilde{\lambda}_3^{(1)} = {2 \dot{\phi}_0 \over \eftcut^2}.
}
Requiring a controlled derivative expansion forces $(\dot{\phi}_0)^{1/2} < \eftcut$.
The quadratic mixing in~\eqref{eq:SE_eft} contributes to the power spectrum $\Delta P_\zeta / P_\zeta \sim \dot{\phi}_0^2/(H^2\eftcut^2)$.
Demanding $\Delta P_\zeta / P_\zeta \lesssim 1$ gives a stronger constraint $\dot{\phi}_0/H \lesssim \eftcut$.
This enforces
\es{eq:SE_eft_con}{
\tilde{\lambda}_2 \tilde{\lambda}_3^{(1)} \lesssim {H^2 \over \dot{\phi}_0} \approx 3\times 10^{-4},
}
where in the last relation we have used the inferred value of the primordial power spectrum. 

\subsubsection{Double Exchange}\label{sec:DoEx}
\begin{figure}[htbp]
    \centering
    \includegraphics[width=0.9\linewidth]{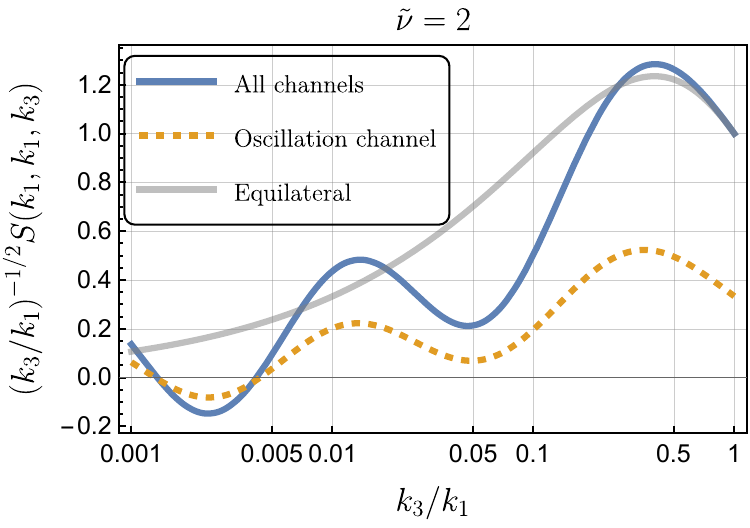}
    \hspace{0.018\linewidth}
    \includegraphics[width=0.9\linewidth]{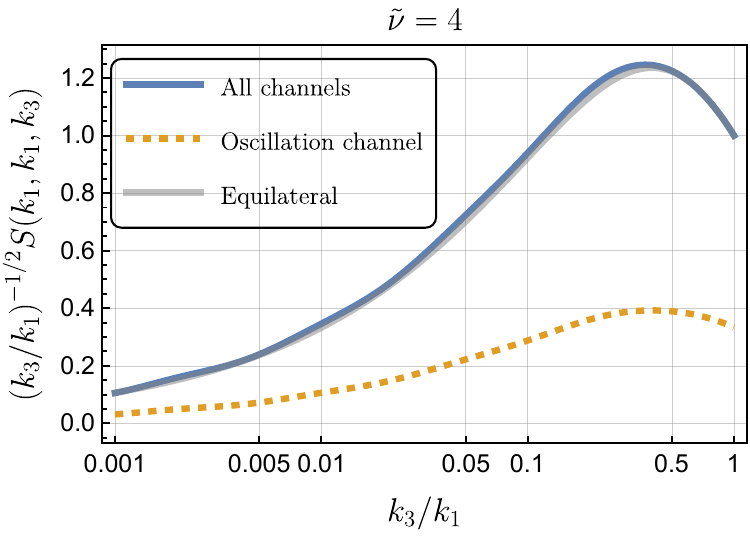}
    \caption{The shape function of double exchange diagrams including all permutations (solid blue) and  after normalizing to unity at equilateral configuration. We have plotted the one oscillation channel (dashed yellow) and equilateral shape (solid gray) for comparison.}
    \label{fig:double_all}
\end{figure}

Now we consider a model for the DE process. The couplings are given as
\begin{equation}\label{DE_model}
    \mathcal{L}_{\rm int}\supset\tilde{\lambda}_2\Lambda a^3\varphi'\sigma+\tilde{\lambda}_3^{(2)}a^3\varphi'\sigma^2.
\end{equation}
Following the same procedure as in the SE model and using~\eqref{cmf_double}, one can construct the shape functions of DE diagrams in the $(k_1,k_2,k_3)$ space. The shapes in isosceles configurations $S(k_1,k_1,k_3)$ of selected masses are shown in Fig.~\ref{fig:double_all}, where we plot one oscillation channel (dashed yellow), where the momentum of $\sigma$ is $k_3$, the sum over all permutations (solid blue), and the standard equilateral shape (solid gray) for comparison. We can see that the oscillations in the full shape are more prominent than the SE case. This is because two among the three permutations contribute to oscillations for DE, while only one does that for SE.

\iffalse
The shape functions in isosceles configurations of selected masses are shown in Fig.~\ref{fig:double_shape}, where we also plot the result given by the bootstrap formulae for comparison. These plots range from squeezed to equilateral configurations, where the bootstrap formulae work well and we can see the agreement of the two methods.
\fi

{\bf Benchmark EFT Interpretation:} The interactions in~\eqref{DE_model} can arise from 
\es{eq:DE_eft}{
&\sqrt{-g}\left({(\partial \phi)^2\sigma\over \eftcut} + {(\partial \phi)^2\sigma^2\over \eftcut^2}\right)\\ &\supset {2\eta\dot{\phi}_0 \varphi'\sigma \over \eta^4\eftcut }  + {2\eta \dot{\phi}_0\varphi'\sigma^2 \over \eta^4\eftcut^2}- {\dot{\phi}_0^2\sigma^2 \over \eta^4\eftcut^2}+\cdots.
}
Comparing with~\eqref{DE_model}, $\tilde{\lambda}_2\Lambda = -2\dot{\phi}_0/\eftcut$, and $\tilde{\lambda}_3^{(2)}=-2\dot{\phi}_0/\eftcut^2$.
The shape function is thus controlled by
\es{}{
(\tilde{\lambda}_2\Lambda)^2 \tilde{\lambda}_3^{(2)} = - {8\dot{\phi}_0^3 \over \eftcut^4}.
}
In this model, there is a `classical' contribution to the mass of $\sigma$, originating from setting both the inflatons to their VEVs in the second term of the LHS of~\eqref{eq:DE_eft}.
Since the oscillatory features are exponentially suppressed for masses larger than Hubble, we require this classical contribution to be smaller than $H^2$, and this enforces $\eftcut > \dot{\phi}_0/H$.
This also ensures the correction to the power spectrum is small.
With this restriction,
\es{eq:DE_eft_con}{
(\tilde{\lambda}_2\Lambda)^2 \tilde{\lambda}_3^{(2)} \lesssim H^4/ \dot{\phi}_0 \approx 3\times 10^{-4} H^2.
}
\subsubsection{Triple Exchange}
\begin{figure}[htbp]
    \centering
    \includegraphics[width=0.9\linewidth]{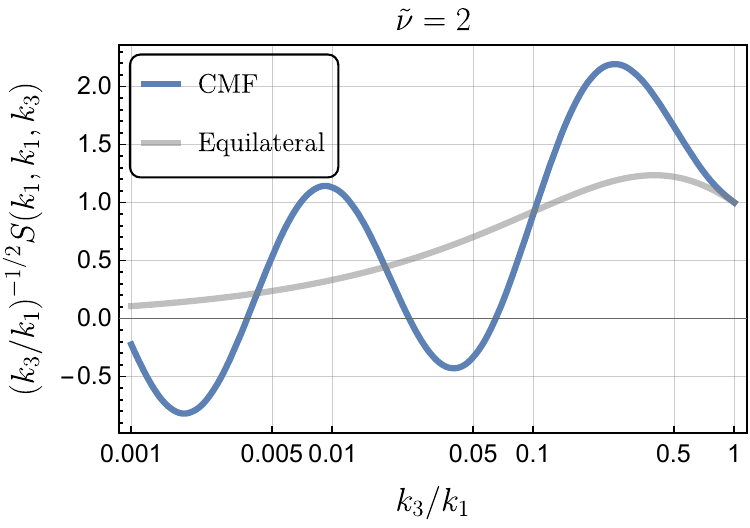}
    \hspace{0.018\linewidth}
    \includegraphics[width=0.9\linewidth]{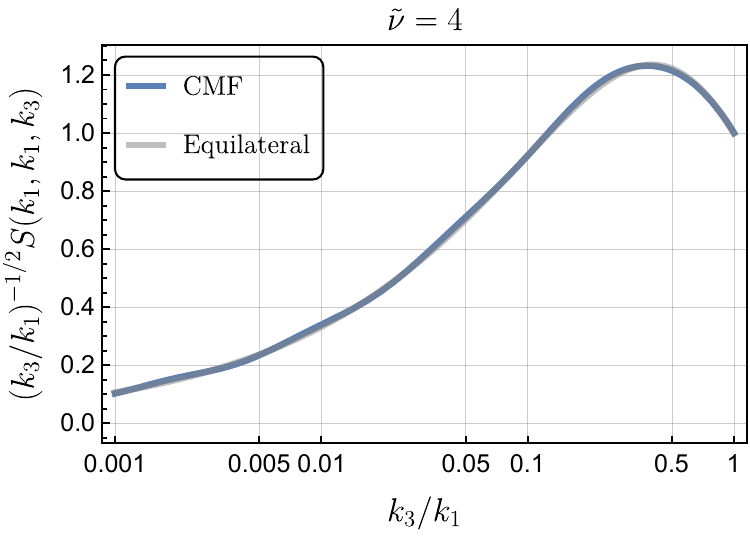}
    \caption{The shapes of triple exchange diagrams for selected masses (solid blue), normalized to unity at equilateral configuration, with equilateral shape (solid gray) for comparison. In this case, the permutation of momenta does not give rise to new contributions.}
    \label{fig:triple_shape}
\end{figure}

We consider the following model for TE,
\begin{equation}\label{TE_model}
    \mathcal{L}_{\rm int}\supset\tilde{\lambda}_2\Lambda a^3\varphi'\sigma+\frac{1}{6}\tilde{\lambda}_3^{(3)}\Lambda a^4\sigma^3.
\end{equation}
Starting from (\ref{cmf_triple}), the procedure is the same as in the previous cases. Moreover, there is only one diagram that contributes to the bispectrum, with no other permutations. The shape functions of isosceles configurations for selected masses are shown in Fig.~\ref{fig:triple_shape}. The oscillatory feature in the shape is much more prominent compared to the SE and DE cases, since in this case taking $k_3\ll k_1$ always correspond to on-shell propagation of $\sigma$.

{\bf Benchmark EFT Interpretation:} The interactions in~\eqref{TE_model} can arise from 
\es{eq:TE_eft}{
\sqrt{-g}\left({(\partial \phi)^2\sigma\over \eftcut} + \kappa \sigma^3\right)\supset {2\eta\dot{\phi}_0 \varphi'\sigma \over \eta^4\eftcut } + {1\over 6}{\kappa \sigma^3\over \eta^4} +\cdots.
}
Comparing with~\eqref{TE_model}, $\tilde{\lambda}_2\Lambda = -2\dot{\phi}_0/\eftcut$, and $\tilde{\lambda}^{(3)}_3\Lambda = \kappa$.
The shape function is controlled by,
\es{eq:TE_eft_con}{
(\tilde{\lambda}_2\Lambda)^3\tilde{\lambda}^{(3)}_3\Lambda  = {8 \dot{\phi}_0^3 \over \eftcut^3}\kappa \lesssim \kappa H^3 \lesssim 4\sqrt{\pi} m_{\sigma}H^3.
}
Here the second-to-last relation follows from requiring the correction to the power spectrum from quadratic mixing to be small, and the last relation from partial wave unitarity constraint on $\kappa$. Importantly, in units of $H$, the constraint in~\eqref{eq:TE_eft_con} is significantly weaker than~\eqref{eq:SE_eft_con} and~\eqref{eq:DE_eft_con}.
This implies the TE diagram can naturally be parametrically larger than the other two, making it a promising target at the cosmological collider.

\subsubsection{Chemical Potential}
For the scenarios described above, the non-analytic signal mediated by the heavy scalar become exponentially suppressed in the $M\gg H$ regime, and instead the NG is dominated by `EFT' contributions peaking in the equilateral configurations.
Therefore, to explore the non-analytic signatures of $M\gg H$ particles, we consider mechanisms where the exponential suppression is offset by some additional sources of energy injection into the system.
To be concrete, we consider the scalar chemical potential (SCP) model described in~\cite{Bodas:2020yho}, consisting of the inflaton and a charged scalar $\chi$ with a softly broken $U(1)$ symmetry.
The Lagrangian is given by,
\es{}{
{\cal L} &= \sqrt{-g}\left(-{1\over 2}(\partial\phi)^2 -V(\phi)-|\partial\chi|^2-M^2|\chi|^2\right.\\
&\left.+ \mt^3(\chi+\chi^\dagger)-{\ii\partial_\mu\phi \over \Lambda}J^\mu-c{(\partial\phi)^2\over \Lambda^2}|\chi|^2\right),
}
where $J^\mu = \chi\partial^\mu\chi^\dagger-\chi^\dagger\partial^\mu\chi$ is the $U(1)$ current.
The parameter $\tilde{m}$ characterizes the soft breaking and $\Lambda$ is an EFT cutoff scale.

The $\chi$ equation of motion is
\es{}{
-\Box \chi - \frac{\ii}{\Lambda} (\Box\phi)\chi - \frac{2i}{\Lambda}\partial^{\mu}\phi\partial_{\mu}\chi \\= -M^2 \chi + \tilde{m}^3 - \frac{c}{\Lambda^2}(\partial_{\mu}\phi)^2 \chi
}
To analyze the dynamics, we decompose the inflaton field around its (slow-roll) classical trajectory $\phi(t,\vec{x}) = \phi_0(t) +\varphi(t,\vec{x}) \approx \dot{\phi}_0t+\varphi(t,\vec{x}) \equiv \omega \Lambda t+\varphi(t,\vec{x})$. We see that the soft-breaking term induces a time-independent vacuum expectation value for the $\chi$ field,
\es{}{
\chi_0 = \frac{-\mt^3}{c\omega^2 - M^2+\frac{\ii}{\Lambda}\Box \phi_0},
}
where we have denoted $\dot{\phi}_0/\Lambda = \omega$.
Expanding the action around this time-independent VEV, $\chi = \chi_0 + \delta\chi$, the Lagrangian involving the $\delta\chi$ perturbation is 
\es{}{
&\frac{\mathcal{L}_{\delta\chi}}{\sqrt{-g}}
%= - \left|\partial_{\mu}\delta\chi\right|^2 -M^2 (\chi_0\delta\chi^\dagger+\chi_0^\dagger\delta\chi+|\delta\chi|^2) +m^3(\delta\chi+\delta\chi^\dagger)
%\\&+\frac{i}{\Lambda}\partial_{\mu}\phi \left((\chi_0^{\dagger}+\delta\chi^{\dagger})\partial^{\mu}\delta\chi-(\chi_0+\delta\chi)\partial^{\mu}\delta\chi^{\dagger}\right)-\frac{c}{\Lambda^2}(\partial_{\mu}\phi)^2 \left(\chi_0\delta\chi^{\dagger}+\chi_0^{\dagger}\delta\chi+|\delta\chi|^2\right)\\
= - \left|\partial_\mu\delta\chi\right|^2 -M^2 (\chi_0\delta\chi^\dagger+\chi_0^\dagger\delta\chi+|\delta\chi|^2)\\ &+m^3(\delta\chi+\delta\chi^\dagger)-\frac{\ii}{\Lambda}\Box\phi \left(\chi_0^{\dagger}\delta\chi-\chi_0\delta\chi^{\dagger}\right)\\& + \frac{\ii}{\Lambda}\partial_{\mu}\phi\left(\delta\chi^{\dagger}\partial^{\mu}\delta\chi - \delta\chi\partial^{\mu}\delta\chi^{\dagger}\right)\\&-\frac{c}{\Lambda^2}(\partial_{\mu}\phi)^2 \left(\chi_0\delta\chi^{\dagger}+\chi_0^{\dagger}\delta\chi+|\delta\chi|^2\right).
}
Using the EOMs for $\chi_0$ and $\chi_0^\dagger$,
\es{}{
&\frac{\mathcal{L}_{\delta\chi}}{\sqrt{-g}}= - \left|\partial_\mu\delta\chi\right|^2 -M^2 |\delta\chi|^2 -\frac{\ii}{\Lambda}\Box\varphi \left(\chi_0^{\dagger}\delta\chi-\chi_0\delta\chi^{\dagger}\right) \\&+ \frac{\ii}{\Lambda}\partial_{\mu}\phi\left(\delta\chi^{\dagger}\partial^{\mu}\delta\chi - \delta\chi\partial^{\mu}\delta\chi^{\dagger}\right)\\&-\frac{c}{\Lambda^2}\left(-2\dot{\phi}_0\dot{\varphi}+\left(\partial\varphi\right)^2\right) \left(\chi_0\delta\chi^{\dagger}+\chi_0^{\dagger}\delta\chi+|\delta\chi|^2\right).
}
Then we perform a phase rotation~\cite{Bodas:2025vpb} on the perturbation $\delta\chi\rightarrow \delta\chi e^{-\ii\phi/\Lambda}$ to remove the $\partial_{\mu}\phi J^{\mu}[\delta\chi]$ term and use $\Box\varphi = 0$ for external inflaton legs, to get
\es{}{
&\frac{\mathcal{L}_{\delta\chi}}{\sqrt{-g}} \supset - \left|\partial_{\mu}\delta\chi\right|^2 -M^2 |\delta\chi|^2 \\&-\frac{c}{\Lambda^2}\left(-2\dot{\phi}_0\dot{\varphi}+\left(\partial\varphi\right)^2\right)\left(\chi_0^{\dagger}\delta\chi e^{-\ii\phi/\Lambda}+{\rm h.c.}\right).
}
Here, we have only kept terms linear in $\delta\chi$ and $\delta\chi^\dagger$ since we are interested in tree-level NG.
This form maintains the shift symmetry of the inflaton since any shift $\phi\rightarrow\phi+{\rm constant}$ either keeps the Lagrangian terms invariant or can be absorbed by doing a global rotation of $\chi$.
The interaction vertices are given by
\es{eq:chem_pot_int}
{
&\text{quadratic mixing:}\;\; \frac{2c\omega}{\Lambda}\dot{\varphi}\chi_0^{\dagger}\delta\chi e^{-\ii \omega t}+\hc,\\
&\text{cubic interaction:}\;\;-\frac{\ii 2 c \omega}{\Lambda^2}\dot{\varphi}\varphi\chi_0^{\dagger}\delta\chi e^{-\ii \omega t}\\&-\frac{c}{\Lambda^2}(\partial\varphi)^2\chi_0^{\dagger}\delta\chi e^{-\ii \omega t}+\hc.
}
So there are two types of cubic interactions, one involving only the time derivative of $\varphi$ and another including covariant derivatives.

\paragraph{Time Derivative Contribution}

Schematically, the three-point function can be written as,
\es{eq:3pt_single_schematic}{
&\langle \varphi(\vec{k}_1)\varphi(\vec{k}_2)\varphi(\vec{k}_3)\rangle \\
&= \sum_{a,b=\pm} (-ab) \int_{-\infty}^0 {\D \eta_1 \over (-\eta_1)^4} \int_{-\infty}^0 {\D \eta_2 \over (-\eta_2)^4}[{\cal O}_1 G_a(\eta_1, k_1)]\\&\times [{\cal O}_2 G_a(\eta_1, k_2)] [{\cal O}_3 G_b(\eta_2, k_3)] \\& \times  D_{ab}(\eta_1, \eta_2, k_3)V_1(\eta_1)V_2(\eta_2) + (k_1\leftrightarrow k_2)+{\text{cyclic perms.}}.
}
%where the operators ${\cal O}$ depend on how $\varphi$ appears in the vertices, and $V_i$ are vertices containing various model-dependent factors.
% The indices $a$ and $b$ denote whether the vertices come time-ordering ($+$) or anti-time ordering ($-$), and the overall minus sign is the $i^2$ factor from the time evolution operator.
% The inflaton bulk-to-boundary propagators are given by,
% \es{}{
% G_a(\eta, k) = {1\over 2k^3} (1-\ii a k \eta)e^{\ii ak \eta},~\partial_\eta G_a(\eta, k) = {\eta\over 2k} e^{\ii ak \eta}.}
Given the time-derivative cubic term, we have
\es{}
{
\mathcal{O}_1 G_a(\eta_1, k_1) = -\frac{\eta^2_1}{2k_1}e^{\ii a k_1 \eta_1},\;\; \mathcal{O}_2 = \mathds{1},\;\;\\ \mathcal{O}_3G_b(\eta_2,k_3) = -\frac{\eta^2_2}{2k_3}e^{\ii b k_3 \eta_2}.}
The mixing vertex correspond to $V_2(\eta) = 2c\omega\chi_0(-\eta)^{-\ii\omega}/\Lambda$ or its complex conjugate, while the cubic vertex is given by $V_1(\eta) = (-\ii)2c\omega\chi_0^\dagger(-\eta)^{+\ii\omega}/\Lambda^2$ or the complex conjugate. Massaging the expression to match the form of the seed integral \eqref{seed}, we obtain
\es{eq:CPtime}{
&\langle \varphi(\vec{k}_1)\varphi(\vec{k}_2)\varphi(\vec{k}_3)\rangle =  {\ii \over 8 k_1 k_2 k_3}\left(\frac{2c\omega}{\Lambda}\right)^2 \frac{|\chi_0|^2}{\Lambda} \\
&\times \sum_{a,b=\pm} \left[\left(\frac{1}{k_2^2k_3}+\frac{1}{k_1^2 k_3}\right)\mathcal{I}_{ab}^{-2-\ii \omega, p_2}(u,1)\right.\\ &\left.+\ii a \left(\frac{1}{k_2 k_3^2}+\frac{1}{k_1 k_3^2}\right)\mathcal{I}_{ab}^{-1-\ii \omega, p_2}(u,1)-(\omega\rightarrow -\omega)\right]\\&+{\text{cyclic perms.}},
}
with $p_2 = -2+\ii\omega$.

\paragraph{Covariant Derivative Contribution.} There is another contribution to the three point function at tree-level from the other cubic interaction.
%For this purpose, we write
%\es{}{
%\tchi_0 = {\tilde{m}^3 \over \left((M_{\rm eff}^2-\omega^2)^2 + 9\omega^2\right)^{1/2}}%e^{i\alpha} (-\eta)^{-i\omega}\equiv {\cal A}e^{i\alpha} (-\eta)^{-i\omega},}
%where $\alpha$ is a constant phase that will drop out of the final three point function.
Repeating the same steps as above, we arrive at the three point function,
\es{eq:cp_shape}{
&\langle \varphi(\vec{k}_1)\varphi(\vec{k}_2)\varphi(\vec{k}_3)\rangle = -{1\over 8 k_1k_2k_3^4}\left(\frac{2c}{\Lambda}\right)^2\frac{\omega|\chi_0|^2}{\Lambda}\times\\
&\sum_{a,b=\pm}\left[\left({\cal I}_{ab}^{-\ii\omega,p_2}+{k_3^2\left(\vec{k}_1\cdot\vec{k}_2\right)\over k_1^2k_2^2}\left({\cal I}_{ab}^{-2-\ii\omega,p_2}\right.\right.\right.\\& \left.\left.\left.+\ii a{k_{12}\over k_3}{\cal I}_{ab}^{-1-\ii\omega,p_2}-{k_1k_2\over k_3^2}{\cal I}_{ab}^{-\ii\omega,p_2}\right)\right)+(\omega\rightarrow-\omega)\right]\\
&+{\text{cyclic perms.}},
}
where we have suppressed the $(u,1)$ argument of the ${\cal I}_{ab}$ function. 

\begin{figure}[htbp]
    \centering
    \includegraphics[width=0.9\linewidth]{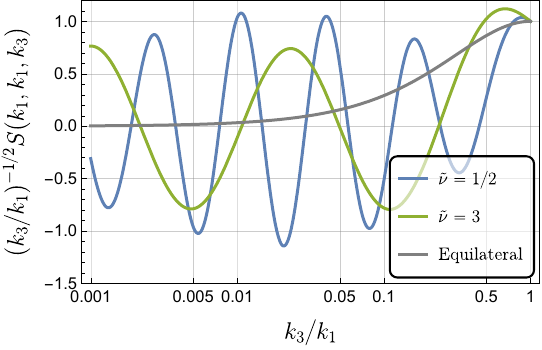}
    \hspace{0.018\linewidth}
    \includegraphics[width=0.9\linewidth]{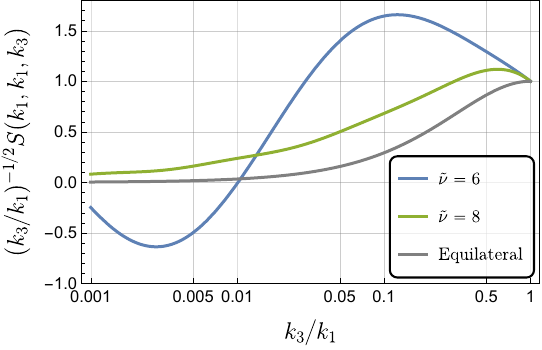}
    \caption{The shape function in the scalar chemical potential model for selected masses and $\omega = 5$, normalized to unity at equilateral configuration, with equilateral shape (solid gray) for comparison. We see that the chemical potential effectively overcomes the Boltzmann suppression and the oscillatory feature is prominent for $\tilde{\nu}<\omega$, with an oscillation frequency $\approx |\omega-\tilde{\nu}|$.}
    \label{fig:cp_shape}
\end{figure}

% In the rest of the paper $\omega$ is always measured in units of $H$. 
The total shape functions, from the time and covariant derivative contributions, in isosceles configurations and for $\omega = 5$ and selected masses are shown in Fig.~\ref{fig:cp_shape}. The energy injection from the chemical potential overcomes the Boltzmann suppression and the oscillatory feature is prominent for $\tilde{\nu}\lesssim \omega$. Moreover, the oscillation frequency is given by the difference $|\omega-\tilde{\nu}|$. At $\tilde{\nu}> \omega$, the shape function approaches a smooth power-law function similar to the standard equilateral shape, but it is not identical to equilateral as for the multiple exchange cases due to the time-dependent coupling between the inflaton and the heavy scalar $\chi$.

\paragraph{Correction to the Power Spectrum.} The quadratic mixing vertex contributes to the power spectrum.
Following similar steps as above we get,
\es{eq:CP_powerspec}{
\langle \varphi(\vec{k}_1)\varphi(\vec{k}_2)\rangle &= \left({2 c\omega \over \Lambda}\right)^2 |\chi_0|^2{1\over 4k_1^3}\\&\times \sum_{a,b=\pm}\left[{\cal I}_{ab}^{-2-\ii\omega,-2+\ii\omega}(1,1)+ (\omega\rightarrow -\omega)\right]\\
&\equiv \left({2 c\omega \over \Lambda}\right)^2 |\chi_0|^2{1\over 4k_1^3}\tilde{\mathcal{I}}(\omega, \nu).
}
The dimensionless function $\tilde{\mathcal{I}}(\omega, \nu)$ ranges from $\mathcal{O}(1)$ to $\mathcal{O}(0.01)$ for the values of $\omega$ and $\tilde{\nu}$ considered in this work. Demanding that this is a subleading contribution to the power spectrum, we get
\begin{equation}
    \left({2 c\omega \over \Lambda}\right)^2 \frac{|\chi_0|^2}{H^2}\tilde{\mathcal{I}}(\omega, \nu) \lesssim 1.\label{eq:CP_constraint}
\end{equation}

\section{Results from the Planck Data}\label{sec:Planck}

We now use the \texttt{CMB-BEST} code~\cite{Sohn:2023fte} to search for the bispectra shapes that we computed above. We first discuss the conventions for the bispectra templates and the numerical procedure used to obtain those.

\paragraph{Normalization}The \texttt{CMB-BEST} pipeline extracts the estimated amplitude $f_{\rm NL}$ from the \textit{Planck} 2018 data, given a dimensionless shape function $\tilde{S}^{\Phi}$ that specifies the bispectrum of the Bardeen potential. Following \textit{Planck} 2018 convention \cite{Planck:2019kim} (and thus \texttt{CMB-BEST}), but adjusting for the fact that all models considered in this paper has exact scale-invariance, we define
\begin{equation}
\begin{split}
    &\expval{\Phi(\vec{k}_1)\Phi(\vec{k}_2)\Phi(\vec{k}_3)}' = B_{\Phi}(k_1, k_2, k_3)\\
    &\equiv \, 6 A^2 f_{\rm NL} \frac{\tilde{S}^{\Phi}(k_1, k_2, k_3)}{k_1^2k_2^2k_3^2},\;\\
    &A = 2\pi^2 \left(\frac{3}{5}\right)^2 A_s,
\end{split}
\end{equation}
where $A_s \equiv (k^3/(2\pi^2)) (k_*/k)^{n_s-1}(5/3)^2\langle\Phi(\Vec{k})\Phi(-\Vec{k})\rangle'$ is the dimensionless power spectrum, and $\tilde{S}^{\Phi}$ is the shape function resulting from the calculation outlined in Sec.~\ref{sec:treefNL}. Note $f_{\rm NL}$ is completely degenerate with an arbitrary normalization convention for $\tilde{S}^{\Phi}$, and a normalization scheme must be chosen to make the value of $f_{\rm NL}$ meaningful. 
A common normalization scheme is $\tilde{S}^{\Phi}(k_*,k_*,k_*)=1$ at some pivot value $k_*$. The standard local, equilateral, and orthogonal shapes are normalized this way. However, this is not a convenient normalization for our models because of the oscillatory nature of the shape functions. Even within the same model, the shape function at the equilateral configuration could change rapidly as we slightly change the value of model parameters such as the scalar mass. This would make the $f_{\rm NL}$ constraints artificially fluctuate, making it difficult to interpret and compare across different parameter values.

%Although a certain model may not correspond to the correct underlying physics, one can nevertheless obtain a best-fit value of $f_{\rm{NL}}$ probably with a large uncertainty. If a certain model correctly describes the underlying physics, we expect that the estimated $f_{\rm{NL}}$ has a small uncertainty $\sigma$, \textit{i.e.} the value of $f_{\rm{NL}}/\sigma$ is large. However, to confirm the correctness of a model, it is necessary to conduct joint analysis with different models, which is not the main purpose of this section.

One possible alternative choice is to not normalize with respect to any particular momentum configuration at all, but to fix all overall multiplicative coupling constants in the model as we change other parameters that affect the shape. For example, in the SE model given by \eqref{SE_model}, we could fix the values of $\tilde{\lambda}_2$, $\tilde{\lambda}_3^{(1)}$, and $\Lambda$, and let the mass parameter $\tilde{\nu}$ vary. By definition, a small change of $\tilde{\nu}$ will only lead to small change in both the amplitude and shape of the calculated bispectrum, and no sudden artificial fluctuation in $\fnl$ constraint would occur. However, an issue with this approach is that the parameters that affect the shape often also affect the amplitude. For example, for SE model, at high-mass limit, the value of the diagram is suppressed by a $1/\tilde{\nu}^2$ factor,\footnote{For DE and TE diagrams, the suppression is $1/\tilde{\nu}^4$ and $1/\tilde{\nu}^6$, respectively.} which is the effective propagator of the heavy field. In that scenario, while the shape becomes almost identical to the equilateral shape, the estimated $f_{\rm{NL}}$ will scale up as $\tilde{\nu}^2$, which is misleading from an observational point of view. It also makes comparison of our results with searches for standard shapes such as the local, equilateral, and orthogonal inconvenient.

Instead, we introduce the normalization in which we fix the maximal value of $|S^{\Phi}|$ in the entire physical momentum region to be 1. To be precise, we define the normalized form of all shape functions resulting from models studied in this paper as
\begin{equation}\label{normalized}
   \tilde{S}^{\Phi}_{\rm model}(k_1, k_2, k_3)\equiv\frac{k_1^2 k_2^2 k_3^2 B_{\tilde{\varphi}}(k_1,k_2,k_3)}{\max \left\{\left|k_1^2 k_2^2 k_3^2 B_{\tilde{\varphi}}(k_1,k_2,k_3)\right|\right\}},
\end{equation}
where the maximum is taken over all inflaton momentum configurations that are physical, and $B_{\tilde{\varphi}}(k_1,k_2,k_3) = \langle \varphi(\vec{k}_1) \varphi(\vec{k}_1) \varphi(\vec{k}_1)\rangle'$ are as calculated in Sec.~\ref{sec:treefNL}. For the standard equilateral and orthogonal shapes, this normalization convention is equivalent to the common $\tilde{S}^{\Phi}(k_*,k_*,k_*) = 1$ normalization.

\paragraph{From $f_{\rm NL}$ to couplings} Because we are computing the bispectrum from full physical models, the estimated $f_{\rm NL}$ from \texttt{CMB-BEST} using the above convention can be translated to values of the coupling constants for each model. Using the method in Sec.~\ref{sec:method}, given some choice of model parameters, we compute the three-point function of the inflaton perturbation in units of Hubble ($\tilde{\varphi}\equiv \varphi/H$),
\begin{equation}
    \expval{\tilde{\varphi}(\vec{k}_1)\tilde{\varphi}(\vec{k}_2)\tilde{\varphi}(\vec{k}_3)}' = \lambda B^{\lambda=1}_{\tilde{\varphi}}(k_1, k_2, k_3),
\end{equation}
where by $\lambda$ we denote the product of all multiplicative coupling constants. For example, for SE, $\lambda = \tilde{\lambda}_2 \tilde{\lambda}_3^{(1)}$, while the dependence on the mass parameter $\tilde{\nu}$ is built into the function $B^{\lambda=1}_{\tilde{\varphi}}$. Using the physical relation between the Bardeen potential and the inflaton perturbation in the superhorizon limit, we have
\begin{equation}
B_{\Phi} = (3/5)^3\lambda B_{\tilde{\varphi}}^{\lambda=1}(2\pi)^3 A_s^{3/2}.
\end{equation}
Combining this with the convention for $f_{\rm NL}$ defined above, we obtain the multiplicative constants $\lambda$, for a given value of estimated $f_{\rm NL}$ from \texttt{CMB-BEST},
\begin{equation}
\lambda(f_{\rm NL}) = \frac{9}{5} \pi f_{\rm NL}A_s^{1/2}\frac{1}{\max\left(\left| k_1^2 k_2^2 k_3^2 B^{\lambda=1}_{\tilde{\varphi}}\right|\right)}.
\end{equation}
Again we emphasize that the dependence on non-multiplicative model parameters, such as $\tilde{\nu}$ and $\omega$ (for the chemical potential), is built into the shape function $k_1^2 k_2^2 k_3^2 B^{\lambda=1}_{\tilde{\varphi}}$.

\paragraph{Numerical sampling} The \texttt{CMB-BEST} pipeline admits custom shape function using either an analytic expression $S_{\rm model}^{\Phi}(k_1, k_2, k_3)$ or $S_{\rm model}^{\Phi}$ evaluated on a discrete grid of $(k_1, k_2, k_3)$. Even though the cosmological bootstrap method in principle produces an analytic expression of $S_{\rm model}^{\Phi}(k_1, k_2, k_3)$, the high complexity of the transcendental functions involved means that in practice the shape function still needs to be passed onto \texttt{CMB-BEST} numerically. For all models considered in this paper, regardless of whether the CMF or bootstrap method is used, we sample 216000 points in the $(k_1, k_2, k_3)$ volume between the \texttt{CMB-BEST} dynamical range $10^{-3}\leq k_1/k_{\rm max}, k_2/k_{\rm max}, k_3/k_{\rm max} \leq 1, k_{\rm max} = 0.209\,{\rm Mpc}^{-1}$, with a uniform $\log_{10}$ grid of 60 points in each $k$ direction. Because of scale invariance and cyclic symmetry, this is computationally equivalent to the sampling of 3600 points in the $(k_2/k_1, k_3/k_1)$ plane. We found that increasing the sampling density by a factor of 2 has negligible affect on the \texttt{CMB-BEST} results. We use the Legendre basis with $p_{\rm max} = 30$ in \texttt{CMB-BEST}, which result in less than 10\% convergence error for all models and parameter ranges considered in this paper and less than 2\% error for the highest SNR parameter values of each model that we discuss below.

%In this section, we will compare the results using the normalization of fixed coupling constant and fixed maximum for the single exchange case in Fig. \ref{fig:cmbbest_SE}, and use the latter for the rest of this section. \ql{Consider removing the comparison plot since it has the same shape as the $f_{\rm NL}$-parameter contour.}

% The shape functions used by the CMB-BEST pipeline are defined in terms of the curvature perturbation $\zeta$, while the templates we have calculated are in $\varphi$-gauge. To make conversions between them, we make use of the relation
% \begin{equation}
%     \zeta=-\frac{H}{\dot{\phi}}\varphi,
% \end{equation}
% and the observed value of the power spectrum that is $\mathcal{P}_\zeta=H^4/(2\pi)^2\dot{\phi}^2=2\times10^{-9}$, and we have
% \begin{equation}
%     \expval{\zeta_{\bf k_1}\zeta_{\bf k_2}\zeta_{\bf k_3}}'=-\frac{(2\pi)^3\mathcal{P}_\zeta^{3/2}}{H^3}\expval{\varphi_{\bf k_1}\varphi_{\bf k_2}\varphi_{\bf k_3}}'.
% \end{equation}
% Note that the factor $1/H^3$ is canceled out by the factor $H$ in the mode function of $\varphi$, we can remove the factor $1/H^3$ in the above relation since we have used the convention $H=1$ in the calculation of $\expval{\varphi_{\bf k_1}\varphi_{\bf k_2}\varphi_{\bf k_3}}'$. As a result, the relation between shape functions defined in different gauges reads
% \begin{equation}
%     S_\zeta(k_1,k_2,k_3)=-(2\pi)^3\mathcal{P}_\zeta^{3/2}S_{\varphi}(k_1,k_2,k_3).
% \end{equation}

\subsection{Single, Double, and Triple Exchanges}
Now we consider the models of weakly-coupled multiple exchanges discussed in \ref{sec:shape}. For all multiple exchange models, we take 50 equally spaced values of the mass parameter in the range $1/10\leq \tilde{\nu}\leq 5$. For each model with a choice of $\tilde{\nu}$, the \texttt{CMB-BEST} pipeline gives a best-fit $f_{\rm{NL}}$ with its uncertainty.

\begin{figure}[htbp]
    \centering
    \includegraphics[width=0.8\linewidth]{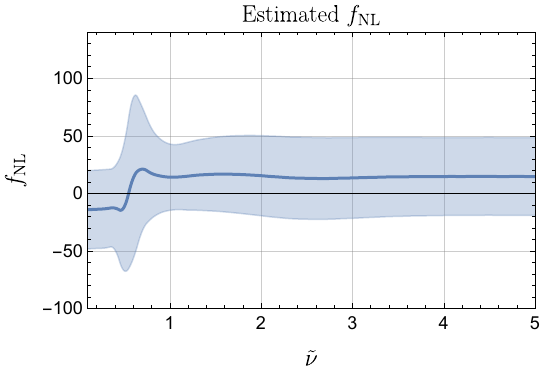}
    \hspace{0.018\linewidth}
    \includegraphics[width=0.78\linewidth]{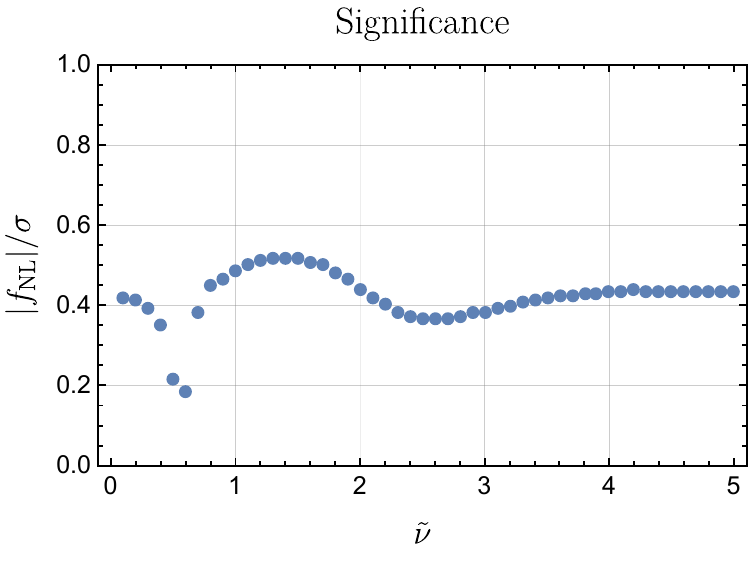}
    \caption{CMB-BEST result for single exchange model. In the top panel, the best-fit values and $1\sigma$ regions are plotted for $\tilde{\nu}$ in the range $0<\tilde{\nu}<5$. In the bottom panel, we plot the significance $|f_{\rm{NL}}|/\sigma$ for each value of $\tilde{\nu}$.}
    \label{fig:cmbbest_SE}
\end{figure}

\paragraph{Single exchange} We plot the estimated $f_{\rm{NL}}$ and the corresponding significance for the SE model in Fig. \ref{fig:cmbbest_SE}. The result becomes constant at large mass $\tilde{\nu}\gtrsim 3$ and the value is almost identical to that of the standard equilateral shape, giving $f_{\rm{NL}}=18\pm42$. The result deviates from the equilateral shape only when $\tilde{\nu}\lesssim1$. The most significant result appears at $\tilde{\nu}=1.4$, with $f_{\rm{NL}}=20\pm39$ and a local significance of 0.52$\sigma$.\footnote{Note that our result for the SE case does not coincide with that given by \cite{Sohn:2024xzd}, because the approximate expression for the shape function used in \cite{Sohn:2024xzd} does not match our exact computation for $\tilde{\nu}\lesssim1$.}

\begin{figure}[htbp]
    \centering
    \includegraphics[width=0.8\linewidth]{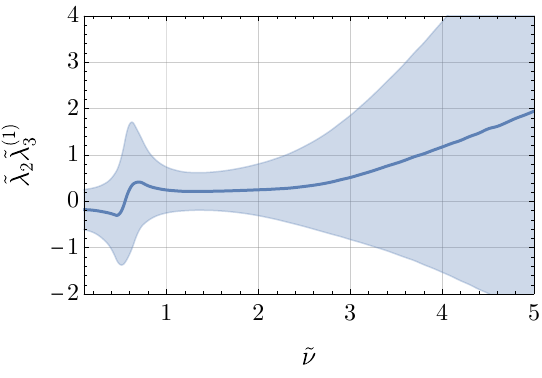}
    \caption{The $f_{\rm NL}$ constraint band in Fig.~\ref{fig:cmbbest_SE} translated to the plane of the dimensionless couplings $\tilde{\lambda}_2\tilde{\lambda}_3^{(1)}$ and $ \tilde{\nu}$. Constraints from perturbativity and power spectrum correction are stronger than this $\fnl$ constraint and are not shown. See text for further details.}
    \label{fig:cmbbest_param_SE}
\end{figure}

To facilitate the physical interpretation of the $f_{\rm NL}$ constraints, in Fig.~\ref{fig:cmbbest_param_SE} we translate the $f_{\rm NL}$ constraint band from Fig.~\ref{fig:cmbbest_SE} to the plane of dimensionless couplings $\tilde{\lambda}_2\tilde{\lambda}_3^{(1)}$ and $\tilde{\nu}$. At large mass, the $f_{\rm NL}$ constraint is constant. However, because of the $1/\tilde{\nu}^2$ suppression from the massive propagator, both the couplings, that correspond to the constant $f_{\rm NL}$ constraint, and their uncertainties grow as $\tilde{\nu}$ increases. As discussed in Sec.~\ref{sec:SiEx}, demanding the interaction between the inflaton and the heavy $\sigma$ to not modify the power spectrum significantly, requires $\tilde{\lambda}_2\tilde{\lambda}_3^{(1)} \lesssim 3\times 10^{-4}$. Comparing this bound to Fig.~\ref{fig:cmbbest_param_SE}, we see that CMB is currently not constraining interesting theory parameter space in this model.

\begin{figure}[htbp]
    \centering
    \includegraphics[width=0.8\linewidth]{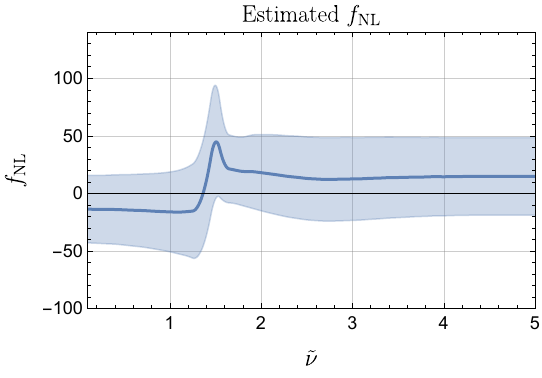}
    \hspace{0.018\linewidth}
    \includegraphics[width=0.78\linewidth]{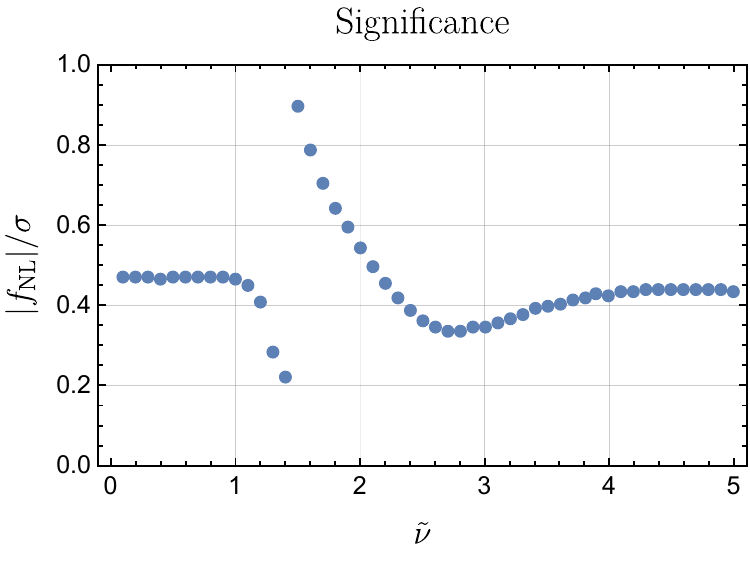}
    \caption{CMB-BEST result of double exchange model. We plotted the best-fit values of $f_{\rm{NL}}$ and $1\sigma$ regions for $0<\tilde{\nu}<5$ in the top panel and the significance $|f_{\rm{NL}}|/\sigma$ in the bottom panel.}
    \label{fig:cmbbest_DE}
\end{figure}

\paragraph{Double exchange} Now we present the result for the DE case, which has not been studied in previous works. The result is shown in Fig. \ref{fig:cmbbest_DE} in the same format as for SE. As in that case, the best-fit value converges to that of the equilateral shape when $\tilde{\nu}\gtrsim3$, giving the correct high-mass limit, but the results at low masses have different features. In particular, the estimated $f_{\rm{NL}}$ has a peak near $\tilde{\nu}=1.5$, with the most significant signal being $f_{\rm{NL}}=55\pm61$ at $\tilde{\nu}=1.5$, with a local significance of 0.90$\sigma$. 

\begin{figure}[htbp]
    \centering
    \includegraphics[width=0.8\linewidth]{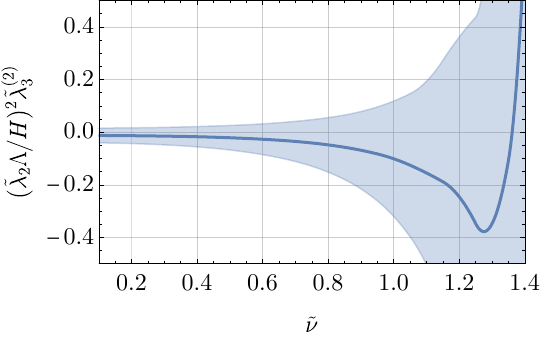}
    \includegraphics[width=0.75\linewidth]{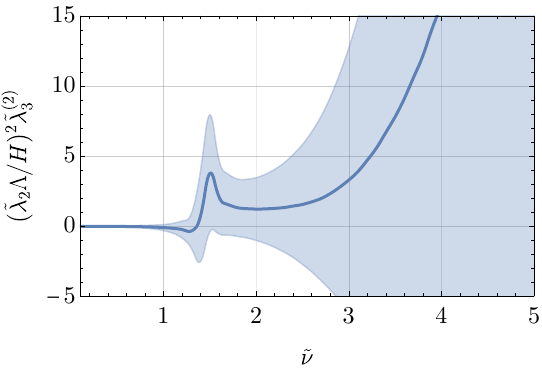}
    \caption{$f_{\rm NL}$ as a function of the dimensionless couplings  $(\tilde{\lambda}_2\Lambda/H)^2\tilde{\lambda}_3^{(2)}$ and $\tilde{\nu}$. The blue line and the band around it correspond, respectively, to the $f_{\rm NL}$ constraint and its uncertainty in Fig.~\ref{fig:cmbbest_DE}. The top panel is a zoomed-in version of the bottom, focusing on the small mass and small coupling regime. Constraints from perturbativity, power spectrum, and correction to the mass of $\sigma$, are stronger than this $\fnl$ constraint and are not shown. See text for further details.}
    \label{fig:cmbbest_param_DE}
\end{figure}

As for the SE model, in Fig.~\ref{fig:cmbbest_param_DE} we plot the value of the dimensionless product $(\tilde{\lambda}_2\Lambda/H)^2\tilde{\lambda}_3^{(2)}$ as a function of $\tilde{\nu}$.
This is translated from the $\fnl$ bounds in Fig.~\ref{fig:cmbbest_DE}. As discussed in Sec.~\ref{sec:DoEx}, demanding that the interaction between the inflaton and the heavy scalar to not modify the power spectrum and an absence of fine-tuning the heavy scalar mass, requires $(\tilde{\lambda}_2\Lambda/H)^2\tilde{\lambda}_3^{(2)} \lesssim 3\times 10^{-4}$. Comparing this bound to Fig.~\ref{fig:cmbbest_param_DE}, we see that CMB is currently not constraining interesting theory parameter space in this model either.

\begin{figure}[htbp]
    \centering
    \includegraphics[width=0.8\linewidth]{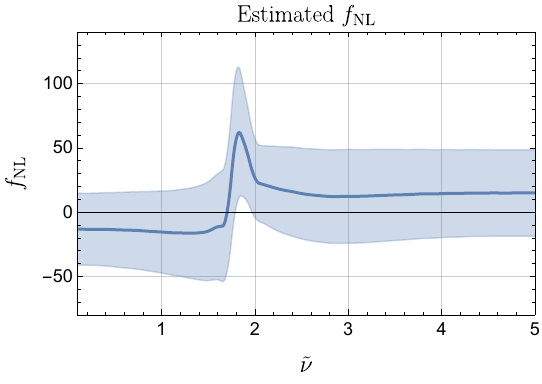}
    \hspace{0.018\linewidth}
    \includegraphics[width=0.8\linewidth]{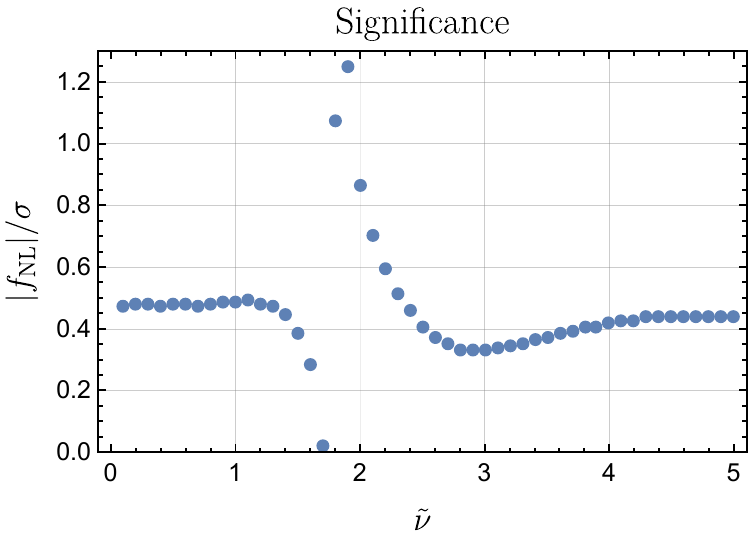}
    \caption{CMB-BEST result of triple exchange model. We plotted the best-fit values of $f_{\rm{NL}}$ and $1\sigma$ regions for $0<\tilde{\nu}<5$ in the top panel and the significance $|f_{\rm{NL}}|/\sigma$ in the bottom panel. Note that the most significant result appears at $\tilde{\nu}=1.9$ with a $1.25\sigma$ significance.}
    \label{fig:cmbbest_TE}
\end{figure}

\paragraph{Triple exchange} The \texttt{CMB-BEST} result for the TE model is shown in Fig. \ref{fig:cmbbest_TE}, which has a similar feature to DE. In this case, the estimated $f_{\rm{NL}}$ has a peak at $\tilde{\nu}=1.8$, with the most significant signal being $f_{\rm{NL}}=62\pm50$ at $\tilde{\nu}=1.9$, with a local significance of 1.25$\sigma$.

\begin{figure}[htbp]
    \centering
    \includegraphics[width=0.8\linewidth]{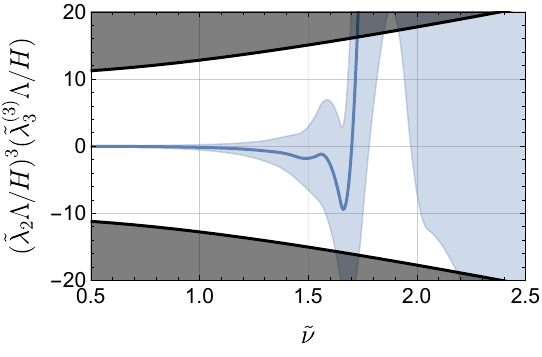}
    \includegraphics[width=0.8\linewidth]{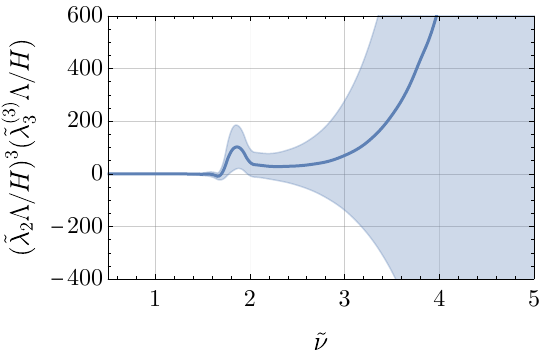}
    \caption{$f_{\rm NL}$ constraint band from Fig.~\ref{fig:cmbbest_TE} translated to the plane of the dimensionless couplings  $(\tilde{\lambda}_2\Lambda/H)^3(\tilde{\lambda}_3^{(3)}\Lambda/H)$ and mass parameter $\tilde{\nu}$. In the top panel, we focus on the small mass regime. In the black shaded region, the correction to the power spectrum is large and the self-interaction of $\sigma$ violates partial wave unitarity bound. In contrast to the SE and DE models, the TE model is non-trivially constrained by the $\fnl$ search for $\tilde{\nu}<2$. In the bottom panel, we extend this to larger $\tilde{\nu}$ and do not show the black shaded region for clarity.}
    \label{fig:cmbbest_param_TE}
\end{figure}

In Fig.~\ref{fig:cmbbest_param_TE} we plot the value of the dimensionless product $(\tilde{\lambda}_2\Lambda/H)^3(\tilde{\lambda}_3^{(3)}\Lambda/H)$, that controls $\fnl$ shown in Fig.~\ref{fig:cmbbest_TE}. Unlike the SE and DE models, where all the couplings that generate the bispectrum lead to a correction to the power spectrum or the heavy scalar mass, in the TE model, $\tilde{\lambda}_3^{(3)}$ is only constrained by a partial wave unitarity bound.
This leads to the much weaker constraint of $(\tilde{\lambda}_2\Lambda/H)^3(\tilde{\lambda}_3^{(3)}\Lambda/H)\lesssim 4\sqrt{\pi} m_{\sigma}/H$. We see that at small mass, bispectrum presents new and stronger constraints on the model parameter space than bounds from perturbativity and correction to the power spectrum.


% \subsubsection{Bump of the constraints of triple exchange diagrams near $\tilde{\nu}=1.8$}
\subsubsection*{A Closer Look at the Triple Exchange Result}
\begin{figure}[htbp]
    \centering
    \includegraphics[width=0.8\linewidth]{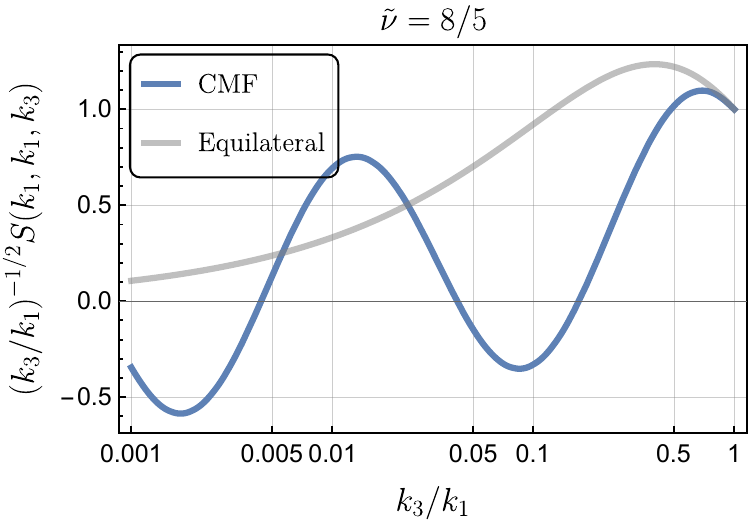}
    \includegraphics[width=0.8\linewidth]{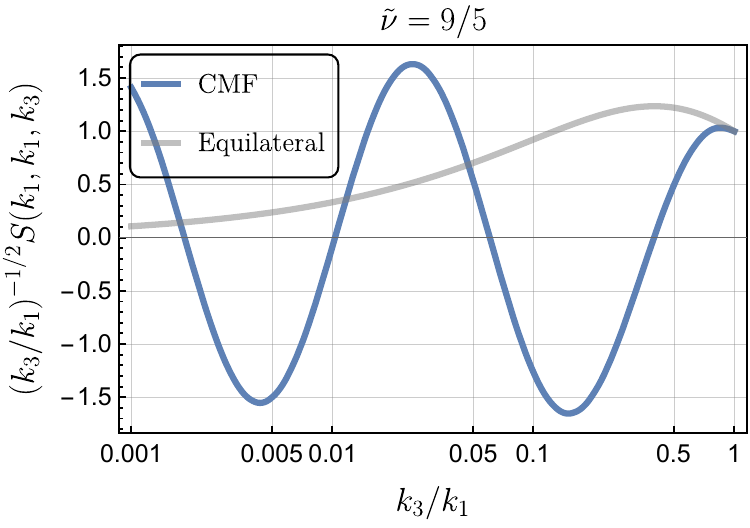}
    \includegraphics[width=0.8\linewidth]{figs/trnu2.pdf}
    \caption{Comparison of shapes of triple exchange diagrams with $\tilde{\nu}=1.6$, $\tilde{\nu}=1.8$ and $\tilde{\nu}=2$. We find that the background part is strongly suppressed at $\tilde{\nu}=1.8$, and for this value the shape function deviates the most from the equilateral shape.}
    \label{fig:shape_TE_comp}
\end{figure}

As can be seen in the top panel of Fig.~\ref{fig:cmbbest_TE}, there is a peak in the constraints near $\tilde{\nu}=1.8$ for TE. Similar peaks with less significance can be observed in the SE and DE cases as well, at smaller values of $\tilde{\nu}$. The peak indicates that the shapes produced at these particular values of $\tilde{\nu}$ have larger deviations from the equilateral shape, and the data prefers these particular shapes more. Indeed, we find that in the TE model, the background part of the shape vanishes at $\tilde{\nu}\approx1.8$, and the full shape is dominated by the oscillatory signal. In Fig.~\ref{fig:shape_TE_comp}, we show the shapes of TE model with $\tilde{\nu}=1.6$, $\tilde{\nu}=1.8$ and $\tilde{\nu}=2$. It can be seen that the background part is strongly suppressed compared to the oscillatory feature at $\tilde{\nu}=1.8$, much more than both $\tilde{\nu}=1.6$ and $\tilde{\nu}=2$. The peak in the $f_{\rm NL}$ constraint provides a hint that the CMB data prefers an oscillatory bispectrum more than a smooth (equilateral) one, although it does not necessarily mean that the TE model correctly describes the underlying physics. We will expand more on this point in the discussion of the chemical potential model in the following subsection.

\subsection{Chemical Potential}\label{sec:chem_pot}

From the multiple exchange models, we see that data slightly prefer the existence of some oscillation in the bispectrum shape. However, the precise frequency that is preferred is difficult to decipher, because in these models a higher frequency of oscillation is always accompanied by an exponential suppression of the oscillation signal.
Thus a scan over large oscillation frequencies is not possible.

In this section, we consider the chemical potential model~\cite{Bodas:2020yho}, where oscillation of a wide range of frequency could be unsuppressed thanks to the energy injection from the rolling of the inflaton. Recall from eqs.~\eqref{eq:CPtime} and~\eqref{eq:cp_shape}, that the chemical potential model has two parameters, $\omega$ that characterize the chemical potential, and hence the energy injection, and the mass parameter $\tilde{\nu}$. 
The oscillation frequency of the shape function is characterized by the difference $|\omega-\tilde{\nu}|$. To investigate whether the data truly prefers the existence of an oscillation, we conduct a comprehensive search in the two-dimensional model parameter space, ranging $\omega$ between 1 and 10 in steps of 1, and $\tilde{\nu}$ in steps of 1/2 between 1/2 and 10. 

\begin{figure}[htbp]
    \centering
    \includegraphics[width=0.9\linewidth]{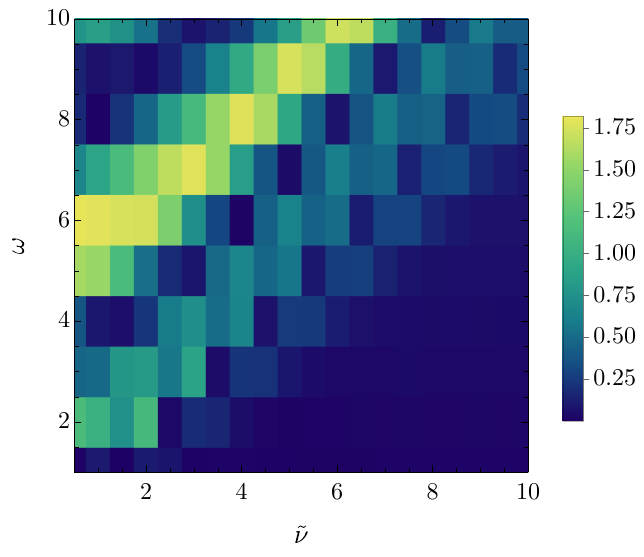}
    \includegraphics[width=0.9\linewidth]{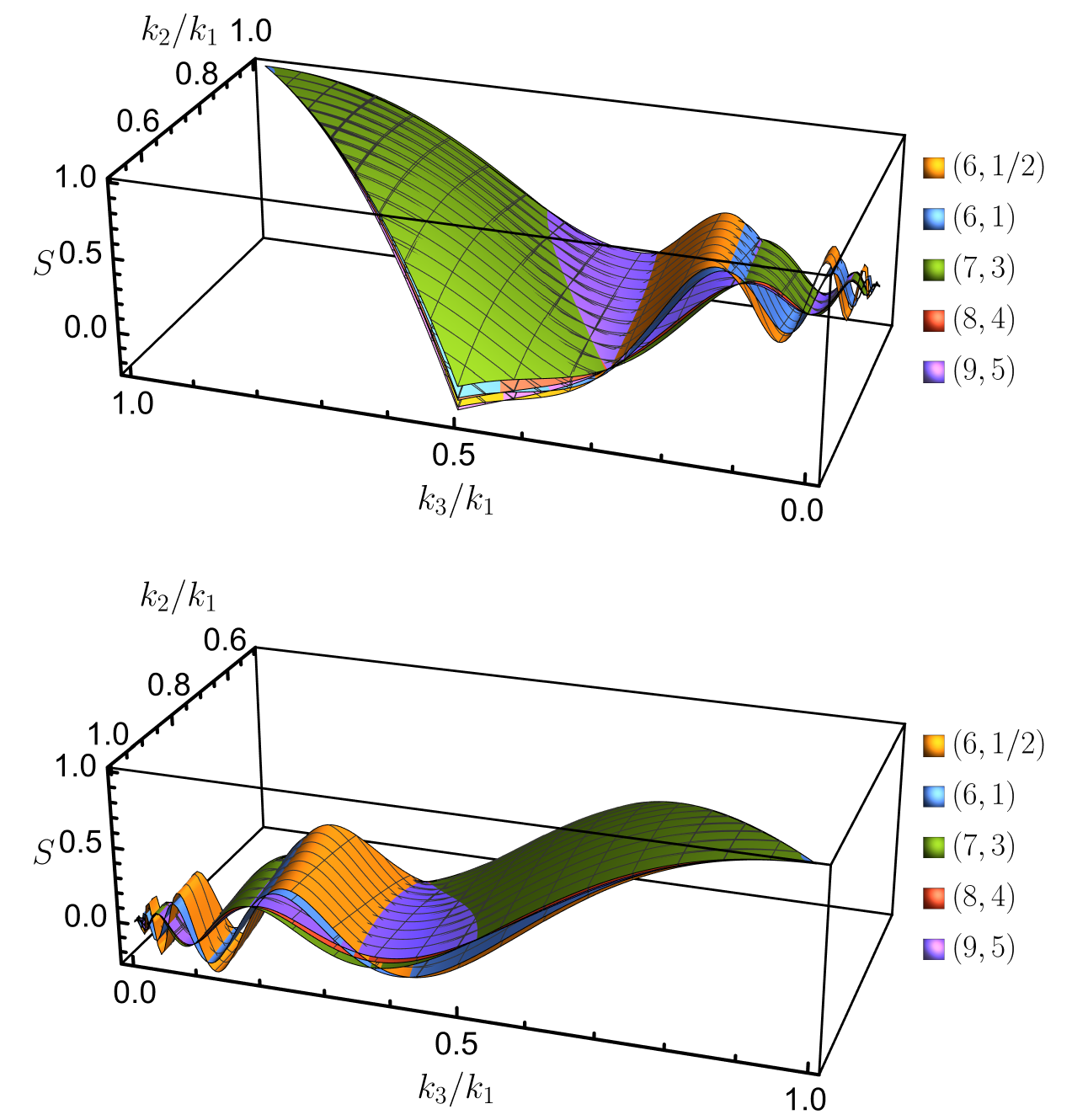}
    \caption{In the top panel, we show the local SNR for the chemical potential model as a function of the oscillation frequency $\omega$ and mass parameter $\tilde{\nu}$. The local SNR is greater than 1.75 for $(\omega, \tilde{\nu}) = (6, 1/2)$, $(6, 1)$, $(7,3)$, $(8, 4)$, and $(9, 5)$. In the middle and bottom panels, we show the normalized bispectrum shape as a function of $k_2$ and $k_3$ with $k_1 = 1$ for these five high-SNR values of $(\omega, \tilde{\nu})$. These two panels show the same shapes from different viewpoints. All five shapes are nearly identical in the equilateral region but not the squeezed region.}
    \label{fig:CPgrid}
\end{figure}

The significance, as a function of $\omega$ and $\tilde{\nu}$, is shown on the top of Fig.~\ref{fig:CPgrid}. The result shows a consistent preference of the CMB data for $\omega - \tilde{\nu} \approx 4$ across a wide range of $\omega$ and $\tilde{\nu}$ values. In particular, the local significance is greater than 1.75$\sigma$ for $(\omega, \tilde{\nu}) = (6, 1/2)$, $(6, 1)$, $(7,3)$, $(8, 4)$, and $(9, 5)$. Using the full correlation matrix from \texttt{CMB-BEST}, we correct for the look-elsewhere effect across the full range of $\omega$ and $\tilde{\nu}$ following~\cite{Sohn:2024xzd}, and find a global significance of 0.7$\sigma$.

On the bottom of Fig.~\ref{fig:CPgrid} we plot the shape function $S(1,k_2,k_3)$ for the five high-SNR points. We see that the five shapes are almost identical in the non-squeezed region $k_3/k_1 \gtrsim 0.3$ while differing more significantly in the squeezed region. More quantitatively, the cosine overlap~\cite{Sohn:2024xzd} between the five high-SNR shapes are above 93\%. Moreover, the cosine overlaps~\cite{Sohn:2024xzd} between the five high SNR chemical potential shapes and the local, equilateral, and orthogonal shape are 17\%, 84\%, and 17\%, respectively; the high correlation with the equilateral shape is consistent with the low global significance of the model.

\begin{figure}[htbp]
    \centering
    \includegraphics[width=0.9\linewidth]{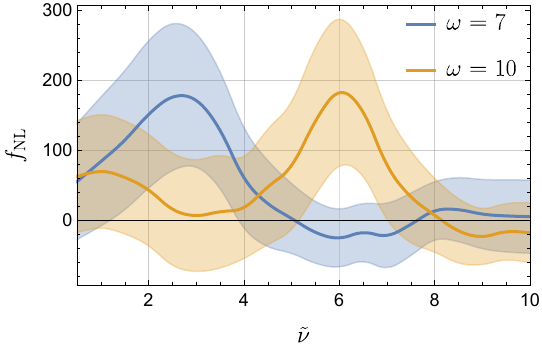}
    \caption{The best-fit values of $f_{\rm{NL}}$ and $1\sigma$ region for the scalar chemical potential model as a function of $\tilde{\nu}$ for $\omega = 7$ and $\omega = 10$. The constraint bands exhibit an approximate shift-invariance for constant $\omega-\tilde{\nu}$ similar to Fig.~\ref{fig:CPgrid}.}
    \label{fig:CP_fnl_om107}
\end{figure}

In Fig.~\ref{fig:CP_fnl_om107} we show the $f_{\rm NL}$ constraint as a function of $\tilde{\nu}$ for two representative choices of $\omega = 7$ and $\omega = 10$. The constraint bands show the same approximate shift-invariance for constant $\omega-\tilde{\nu}$ that we observed from the SNR grid in Fig.~\ref{fig:CPgrid}.  Similar to the multiple exchange model, we can also translate the $f_{\rm NL}$ constraints to model parameters for the scalar chemical potential. Because of the relative $\omega$ factor between the temporal derivative contribution~\eqref{eq:CPtime} and the covariant derivative contribution~\eqref{eq:cp_shape}, $f_{\rm NL}$ is dominated by the temporal contribution when $\omega$ is large, and it's parameterically given by
\begin{equation}
    f_{\rm NL} \sim \frac{\Delta P_{\zeta}}{P_{\zeta}} \frac{\omega {\cal N}}{\tilde{\mathcal{I}}(\omega, \tilde{\nu})}\lesssim \frac{\omega {\cal N}}{\tilde{\mathcal{I}}(\omega, \tilde{\nu})},\label{eq:CP_fnl_size}
\end{equation}
where $\Delta P_{\zeta}/P_{\zeta}$ is the fractional correction to the power spectrum, ${\cal N}={\rm Max}\left\{|k_1^2 k_2^2 k_3^2 B(k_1,k_2,k_3)|\right\}$ is the normalization factor as discussed at the beginning of Sec.~\ref{sec:Planck}, and $\tilde{\mathcal{I}}(\omega, \tilde{\nu})$ characterizes the $\omega$ and $\tilde{\nu}$  dependence of the correction to the power spectrum, as defined in~\eqref{eq:CP_powerspec}. There is an enhancement to the magnitude of $f_{\rm NL}$ at high $\omega$, but even at $\omega = 10$, $\tilde{\mathcal{I}}(\omega = 10, \tilde{\nu})\sim 0.5$ in the range of $\tilde{\nu}$ we searched for. While the normalization ${\cal N}$ varies from $\sim 1$ at $\tilde{\nu} = 1/2$ to $\sim 0.01$ at $\tilde{\nu} = 10$, leading to $f_{\rm NL} \lesssim 20$ to $0.2$. From Fig.~\ref{fig:CP_fnl_om107}, we see that for values of $\nu$ where the constraint is consistent with null within 1$\sigma$, the above constraint from power spectrum correction is currently stronger than the bispectrum search. For the mass values $\tilde{\nu} \in \{5.5,6,6.5\}$, there is a mild local significance (1.8$\sigma$) for nonzero $f_{\rm NL}$ peaking at $\tilde{\nu} = 6$.
The power spectrum analysis above indicates that this significance cannot be consistently attributed to the chemical potential model with the specific parameter values considered here.

But the power spectrum constraint can be evaded in other regions of parameter space. As discussed earlier and is clear from Figs.~\ref{fig:CPgrid} and~\ref{fig:CP_fnl_om107}, the significance with similar central value of $f_{\rm NL}$ exists in a range of $\omega$ and $\tilde{\nu}$ as long as $\omega - \nu \approx 4$. Since the upper bound on the magnitude of $f_{\rm NL}$ grows with $\omega$ from~\eqref{eq:CP_fnl_size}, if for example a similar significance is observed for $\omega = 50$ and $\nu = 47$, the observed $f_{\rm NL}\sim 100$ could be within the upper bound of~\eqref{eq:CP_fnl_size}.

In the full chemical potential model, we have both the time-derivative and covariant derivative cubic interaction between the inflaton and the heavy scalar, contributing to the bispectrum with a fixed relative coefficient of $\omega$, as shown in~\eqref{eq:CPtime} and~\eqref{eq:cp_shape}. But given the mild significance of the full chemical potential model, it is worth treating the time derivative and covariant derivative contribution as independent templates, to see which interaction is driving the significance of the total model. 

\begin{figure}
    \centering
    \includegraphics[width=0.9\linewidth]{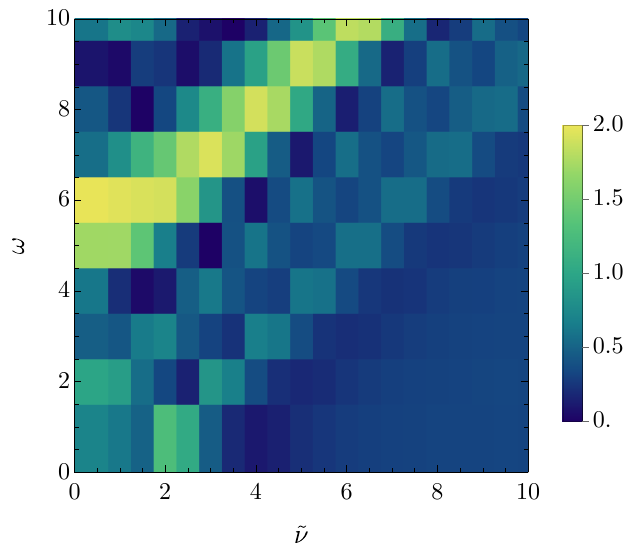}
    \includegraphics[width=0.8\linewidth]{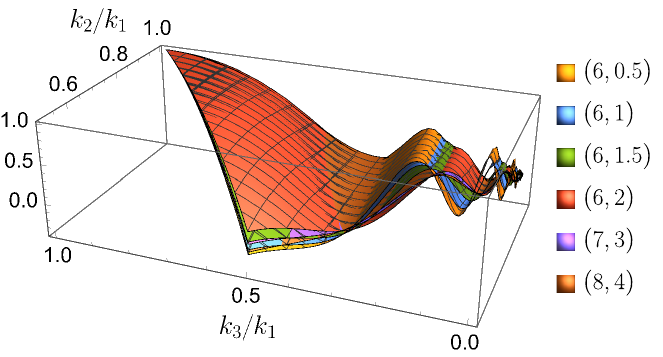}
    \includegraphics[width=0.8\linewidth]{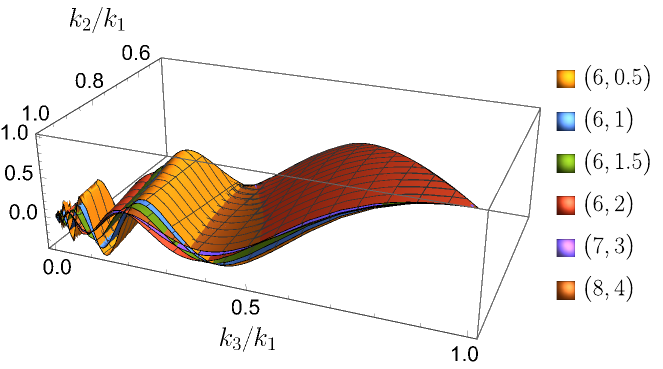}
    \caption{Same as Fig.~\ref{fig:CPgrid}, but for the time derivative component. Compared to Fig.~\ref{fig:CPgrid}, the grid maintains the pattern of the highest significance along the $\omega -\tilde{\nu} \approx 4$ diagonal, but with higher significance than the full model. The local SNR is greater than 1.9 for $(\omega, \tilde{\nu}) = (6, 1/2)$, $(6, 1)$, $(6, 1.5)$, $(6, 2)$,$(7,3)$, and $(8, 4)$. In the middle and bottom panels, we show the normalized bispectrum shape as a function of $k_2$ and $k_3$ with $k_1 = 1$ for these six high-SNR values of $(\omega, \tilde{\nu})$.}
    \label{fig:CPgrid_temp}
\end{figure}

\begin{figure}
    \centering
    \includegraphics[width=0.9\linewidth]{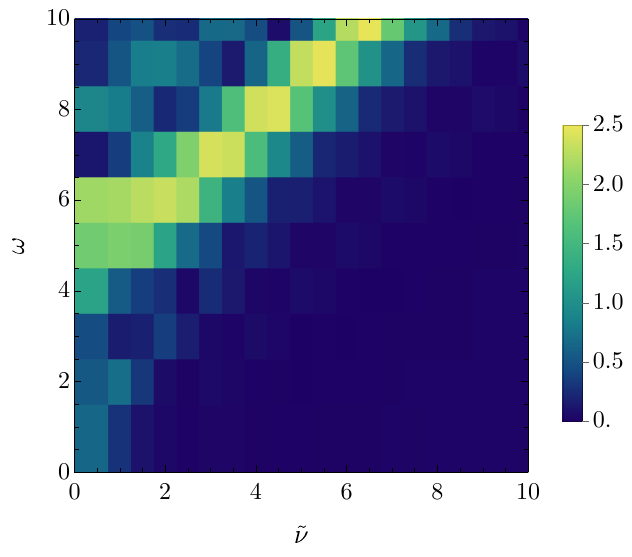}
    \includegraphics[width=0.8\linewidth]{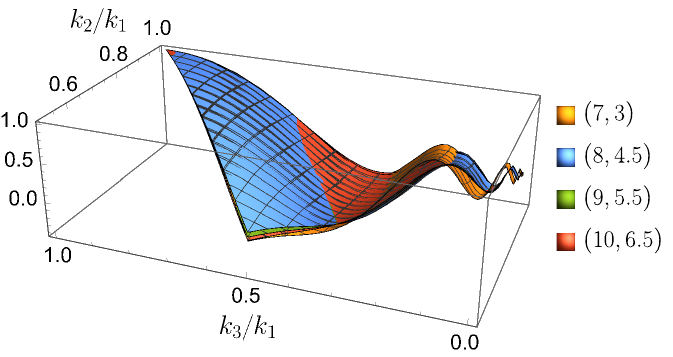}
    \includegraphics[width=0.8\linewidth]{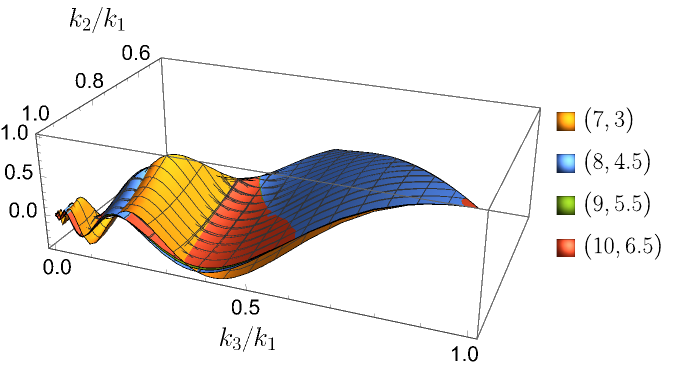}
    \caption{Same as Fig.~\ref{fig:CPgrid}, but for the covariant derivative component. Compared to Fig.~\ref{fig:CPgrid}, the grid has the highest significance along a slightly shifted diagonal $\omega -\tilde{\nu} \approx 3.5$ and with higher significance than the full model. The local SNR is greater than 2.4 for $(\omega, \tilde{\nu}) = (7, 3)$, $(8, 4.5)$, $(9, 5.5)$, and $(10, 6.5)$. In the middle and bottom panels, we show the normalized bispectrum shape as a function of $k_2$ and $k_3$ with $k_1 = 1$ for these four high-SNR values of $(\omega, \tilde{\nu})$.}
    \label{fig:CPgrid_cov}
\end{figure}

In the top panels of Fig.~\ref{fig:CPgrid_temp} and Fig.~\ref{fig:CPgrid_cov}  we show the local SNR as a function of $\omega$ and $\tilde{\nu}$ for the time derivative and covariant derivative contribution. We see that the SNR grid maintains the pattern of having the highest SNR along a constant $\omega-\tilde{\nu}$ diagonal, but the maximum local SNR is higher than the full model: it is $2.0\sigma$ for the time derivative component and $2.5\sigma$ for the covariant derivative. Correcting for look-elsewhere effect as before, we obtain a global 1.0$\sigma$ for the time derivative component and a global 1.5$\sigma$ for the covariant derivative component.

In the middle of bottom panels of Fig.~\ref{fig:CPgrid_temp} and Fig.~\ref{fig:CPgrid_cov} we plot the normalized shape function of the highest-SNR parameter values. Similar to the full model, we again see that the high-SNR shapes are very similar in the non-squeezed region while differing in the squeezed region. This suggests that the significance we observed is dominated by the behavior of the shape function in the non-squeezed region. This is expected since the oscillatory piece of the shape function decays as $(k_3/k_1)^{1/2}$ in the squeezed limit, while the observational errors become larger in the squeezed limit~\cite{Philcox:2026njr}. We emphasize that this does not invalidate the importance of having an enhanced oscillatory feature since it results in a broad peak in the equilateral region that is functionally distinct from power-law dominated shapes, such as the equilateral and orthogonal shapes. The cosine overlap between the local, equilateral, and orthogonal shapes with the high-SNR shapes are below 15\%, 80\%, and 17\% for the temporal component and 5\%, 55\%, and 22\% for the covariant component. As expected, the covariant component that has the highest observed significance has the lowest cosine overlap with the standard shapes, showing that their respective local significances comes from distinct physical origins. On the other hand, for the multiple exchange models with suppressed oscillatory feature, the equilateral region is dominated by equilateral bispectrum-like shape in the large mass regime.

The relative insignificance of the squeezed region could also be heuristically seen from the following. We take the $(\omega, \tilde{\nu}) = (8,4.5)$ template of the covariant component, that has a $2.5\sigma$ local significance, as an example. We then conduct a numerical experiment, searching for this shape template using \texttt{CMB-BEST}, but set the shape template to zero whenever $k_{\min}/k_{\max} < \epsilon$, where $k_{\min}$ and $k_{\max}$ are the minimum and maximum momentum in each $(k_1, k_2, k_3)$ configuration. $\epsilon$ is therefore the squeezedness cutoff, which we vary between $10^{-3}$ and 0.95. We find that as we increase $\epsilon$, the local significance does not change by more than 0.2$\sigma$ from the 2.5$\sigma$ of the original search until $\epsilon = 0.3$, confirming that the most of significance comes from the non-squeezed kinematic region.

%Whether this signifies the need for a more sophisticated data analysis strategy that extract more information from the squeezed region or that the data actually prefers a completely new category of models is beyond the scope of this work.

We reiterate that the bispectrum templates made from the individual components are not physical in the context of the chemical potential model, since they always appear together with a relative coefficient of $\omega$. We present the search result for the individual components to highlight the preference for an oscillatory signature in the data, and to motivate searches for broader categories of cosmological collider models in the CMB, since there very well could be physical models that produce similar bispectrum shapes with even larger SNR.

\section{Discussion}\label{sec:discussion}
In this paper, we computed the full bispectrum shape from the single, double, and triple exchange models and the chemical potential-enhanced production of a heavy scalar.
The triple exchange and chemical potential shapes are calculated for the first time in the literature, while the double exchange shape is calculated for a wider range of scalar masses.
For the multiple exchange models we used the coupled mode function method to numerically compute the bispectrum. For the chemical potential model, the presence of three degrees of freedom (one real inflaton and one complex heavy scalar) make the application of this method more involved.
Therefore, we use the existing general analytic formula of the cosmological bootstrap method.

Using these methods, we constructed the bispectrum shape on a grid in momentum space for each model, and searched for these shapes in the \textit{Planck} CMB data using the \texttt{CMB-BEST} code. We varied the heavy scalar mass between $\tilde{\nu}\in[1/10, 5]$ for multiple exchange, and $\tilde{\nu}\in[1/2, 10]$ and $\omega\in [1, 10]$ for chemical potential.

For multiple exchange, the search resulted in no significant preference, with a local SNR below 1.2$\sigma$ across all mass values and topologies considered. However, for triple exchange, the SNR as a function of $\tilde{\nu}$ show a sudden increase around $\tilde{\nu} = 1.8$, which is where the smooth background contribution to the bispectrum is suppressed compared to the oscillatory piece. The vanishing of the background at $\tilde{\nu} = 1.8$ is simply by numerical coincidence, but it provides a hint that the data has preference for an unsuppressed oscillatory signal.

This preference persists in the searches for the chemical potential model. We find a local SNR of $1.75\sigma$ for the full shape in the chemical potential model at $\omega - \tilde{\nu} \approx 4$.
However, isolating the shape mediated by the covariant derivative coupling leads to a local SNR of $2.5\sigma$ at $\omega - \tilde{\nu} \approx 3.5$, persistent across the full range of $\omega$ and $\tilde{\nu}$ studied. The local SNR is found to be largely insensitive to the squeezed kinematic region, and stays relatively constant even when the shape template is set to zero for $k_{\rm min}/k_{\rm max}$ smaller than 0.3.

Our work demonstrates, from several directions, the importance of constructing the ``complete'' bispectrum shape template from a physical model, where complete refers to two different aspects.
First, it is in the sense of including both the oscillatory ``signal'' and the smooth ``background''. Comparison of \texttt{CMB-BEST} results for the multiple exchange and the chemical potential templates clearly demonstrates that oscillatory templates with or without the presence of a smooth background produces vastly different values of central $f_{\rm NL}$ and SNR. Not accounting for the background contribution from a physical model will thus lead to misleading results. The second aspect of ``completeness'' refers to the fact that we cover the full physical kinematic region, instead of constructing the template from only the squeezed-limit calculation. Intuitively, the non-squeezed region contains greater amount of statistics than the squeezed limit, and therefore it is important that the template is accurate in the non-squeezed region. The importance of this region is confirmed with our numerical experiment on the high SNR covariant component of the chemical potential shape, as discussed before.

While in this work we have made progress in searching for cosmological collider signal in data, many directions of future work remain. 
\begin{itemize}
    \item We have argued and observed that the oscillations in the bispectrum becomes more and more distinct as we go from single exchange to double exchange to triple exchange to chemical potential. Interestingly, our searches indicate that the SNR also keeps increasing in the same way, with the covariant component of the chemical potential template produced the highest global significance of $1.5\sigma$. Is there a complete physical model that results in an even larger and statistically significant SNR?
    \item The search results seem to be insensitive to the squeezed region, essentially seeing only the first period of the oscillatory signal. But the squeezed region is essential to distinguish different templates that currently has the same SNR values (Fig.~\ref{fig:CPgrid} bottom panel). Is it possible to exploit the periodic nature of the oscillatory shape and extract more information from the squeezed region?
    \item As mentioned in the introduction, another known mechanism of enhancing the oscillatory signature of heavy scalar is primordial feature~\cite{Chen:2022vzh}. Primordial feature models are computationally more expensive to construct the shape function in full kinematics because of the lack of continuous scale invariance. But given the hint of oscillatory signature in the {\it Planck} data, it is certainly worth conducting a full kinematic search for primordial feature models as well.
    \item In this work we took the approach of starting from concrete particle models and emphasized the importance of including the background contribution to the bispectrum to accurately search for a given model in the full cosmological data. However, computation of cosmological collider signal from a particle physics model is not only technically difficult but also limited by the range of models being built in the literature. It then bears the question of whether it is possible to ``filter'' the cosmological data directly to search for oscillatory features in a more model-agnostic way.
\end{itemize}

\section*{Acknowledgments}
We thank Arushi Bodas, Xingang Chen, Oliver Philcox, Wuhyun Sohn, Raman Sundrum, and Dong-Gang Wang for useful discussions and comments.
QL is supported in part by the NSF grants PHY-2210498 and PHY-2514611 and by the Simons Foundation. ZX is supported by NSFC under Grants No.\ 12275146 and No.\ 12247103, the National Key R\&D Program of China (2021YFC2203100), and the Dushi Program of Tsinghua University.
This material is based upon work supported by the NSF under grant number 2018149. 
The authors acknowledge the Tufts University High Performance Compute Cluster (https://it.tufts.edu/high-performance-computing) which was utilized for the research reported here.

\appendix

\section{Numerical Performance of Bootstrap and CMF Methods}\label{app_numerical}
In this appendix, we compare the numerical performance of the bootstrap formulae and CMF method when evaluating the SE and DE diagrams. In particular, we highlight the failure of numerical stability of the boostrap formulae near the folded limit. The problem is related to the fact that both the expressions for the signal and background part of the diagram are divergent at the folded limit, and the divergence is to be canceled order by order to reach a finite result. Numerically, the divergence cancellation procedure will inevitably become unstable when the configuration is sufficiently close to the folded limit. In practice, if the numerical instability occurs at the momentum configuration within the momentum range relevant to observation, we have to use the numerical method instead of the analytical formulae for the evaluation of templates. In the following context, the definition of momenta $\mathbf{k}_1$, $\mathbf{k}_2$ and $\mathbf{k}_3$ follows from Fig.~\ref{fig:topo}.

\subsection{Single Exchange}
\begin{figure}[htbp]
    \centering
    \includegraphics[width=0.8\linewidth]{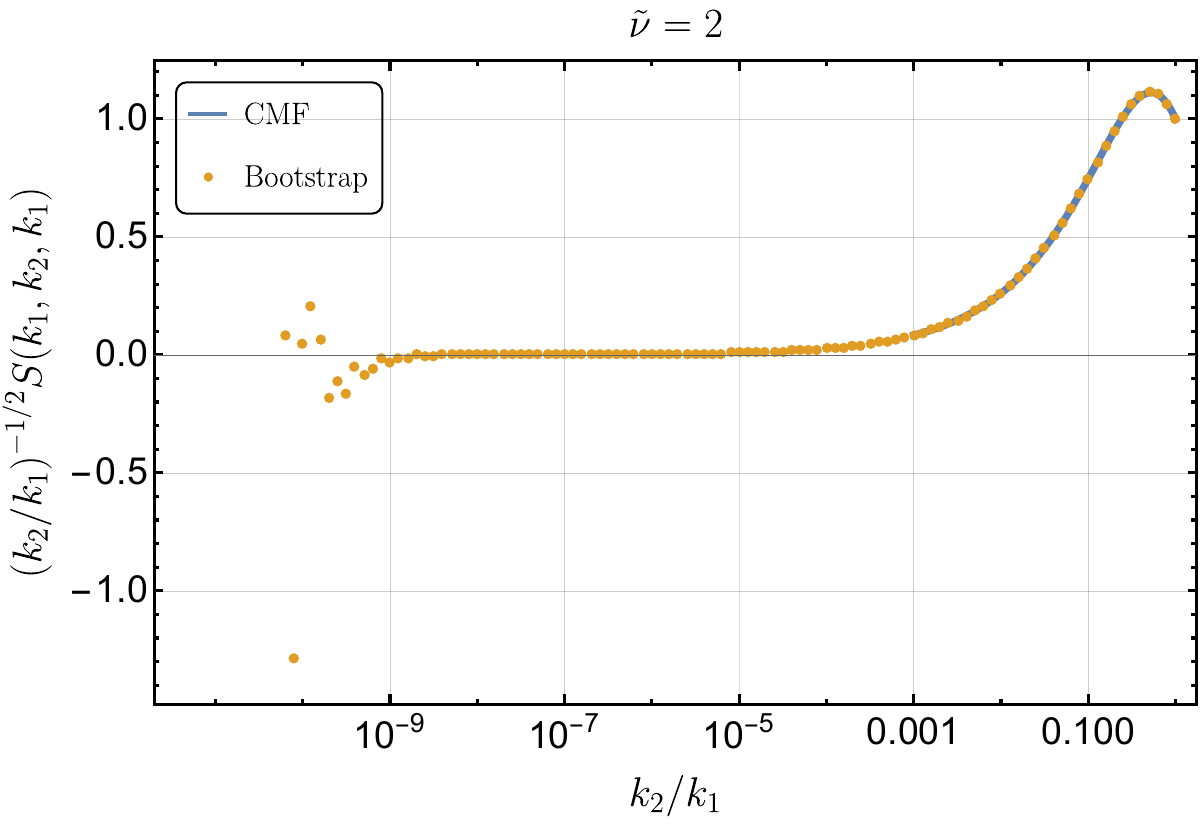}
    \caption{Comparison of CMF and bootstrap result for the permutation of SE diagram with $\tilde{\nu}=2$ that does not contribute to oscillatory signals, \textit{i.e.} the isosceles configurations that ${\bf k}_2$ are squeezed. This configuration corresponds to the folded limit of the bootstrap formulae, where the cancellation of spurious divergence are needed. In the SE case, two methods are in perfect agreement within the kinematic region of interest, \textit{i.e.}, $10^{-3}<k_2/k_1<1$, and the numerical instability occurs at the momentum ratio as small as $k_2/k_1\sim10^{-9}$.}
    \label{fig:single_folded}
\end{figure}

For SE, the folded limit of the bootstrap formulae appears when one of the external lines $\mathbf{k}_1$ and $\mathbf{k}_2$ is squeezed, and $u\equiv 2k_3/(k_1+k_2+k_3)\to1$. To investigate the appearance of numerical instability, we calculate the shape $S(k_1,k_2,k_1)$ produced by the leftmost diagram in Fig. \ref{fig:topo} for $\tilde{\nu}=2$ (as a representative value) as a function of $k_2/k_1$ with both methods, and the result is shown in Fig. \ref{fig:single_folded}. From this figure, we find that the numerical instability of the bootstrap formulae appears when $k_2/k_1$ is as small as $\sim 10^{-9}$, far smaller than momentum ratios probed by the current data. Therefore, we conclude that bootstrap and the CMF method work equally well for SE diagrams.

\begin{figure}[htbp]
    \centering
    \includegraphics[width=0.7\linewidth]{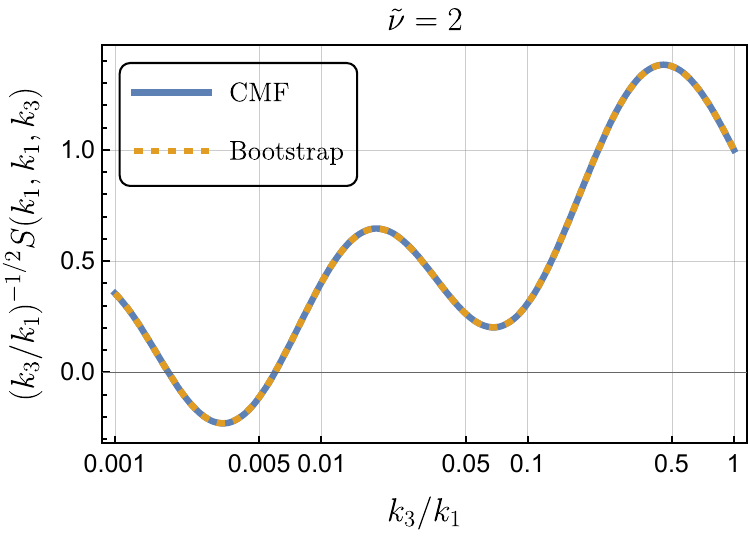}
    % \hspace{0.018\linewidth}
    \includegraphics[width=0.7\linewidth]{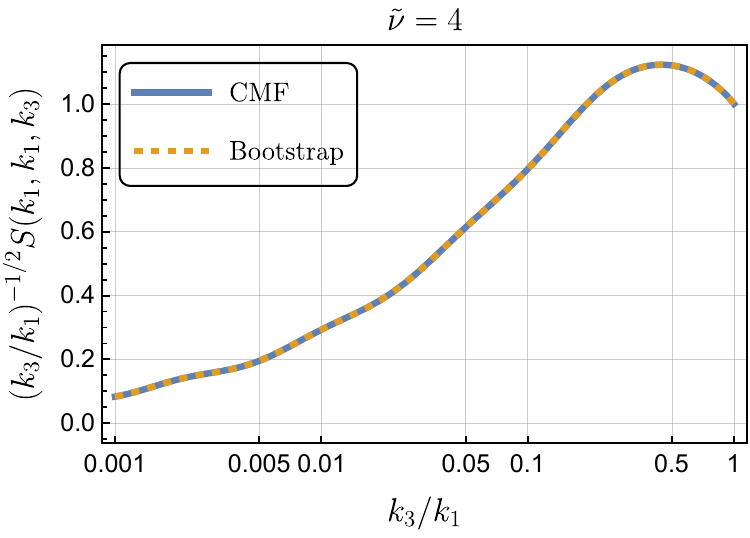}
    \includegraphics[width=0.7\linewidth]{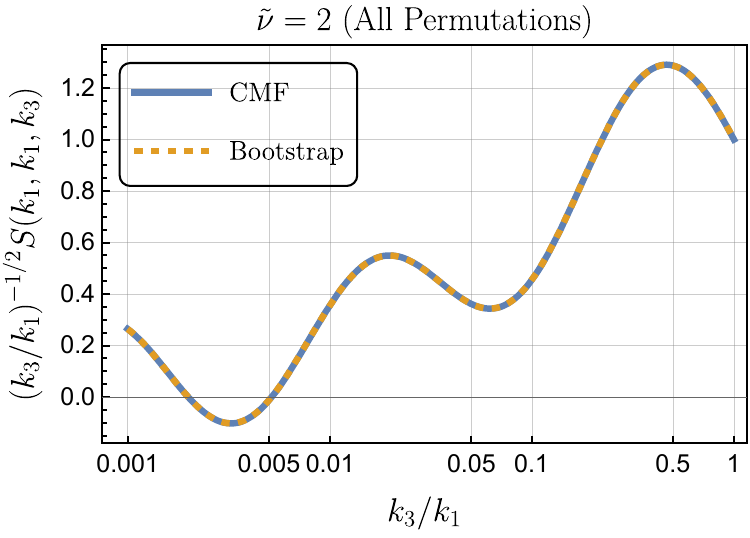}
    % \hspace{0.018\linewidth}
    \includegraphics[width=0.7\linewidth]{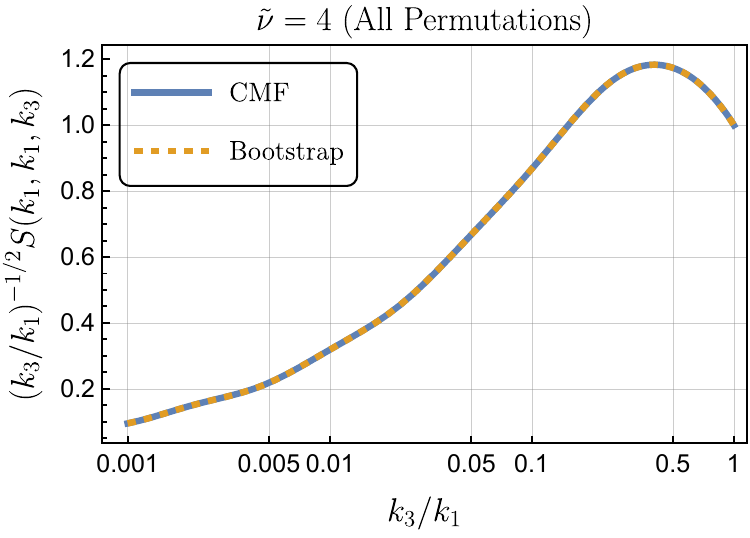}
    \caption{The shapes of SE diagrams in isosceles configurations for $\tilde{\nu}=2$ and $\tilde{\nu}=4$ as two representative values of $\tilde{\nu}$, evaluated by both two methods. The upper panels focus only on the single diagram that produces oscillatory signals in the squeezed limit $k_3\to0$, while other permutations are not included. We have added a prefactor $(k_3/k_1)^{-1/2}$ to the scale invariant shape function $S(k_1,k_1,k_3)$ so that the amplitude of oscillatory signal is constant in the squeezed limit, and the shapes are normalized at the equilateral configuration, i.e., $S(k,k,k)=1$. The lower panels gives the shape functions after summing up all three permutations. In all cases we find the perfect agreement between the CMF and bootstrap methods.}
    \label{fig:single_shape}
\end{figure}

As a verification, we plot the shapes evaluated by both methods for two representative masses $\tilde{\nu}=2$ and $\tilde{\nu}=4$ in Fig. \ref{fig:single_shape}. Indeed, we see that both methods are capable of providing accurate templates of the full shape and the results are perfectly consistent. %In practice, the bootstrap formulae are more flexible since it can be applied to the case of oscillatory couplings in a trivial way, while it takes more effort for the CMF method to do the same thing.

\subsection{Double Exchange}

\begin{figure}[htbp]
    \centering
    \includegraphics[width=0.8\linewidth]{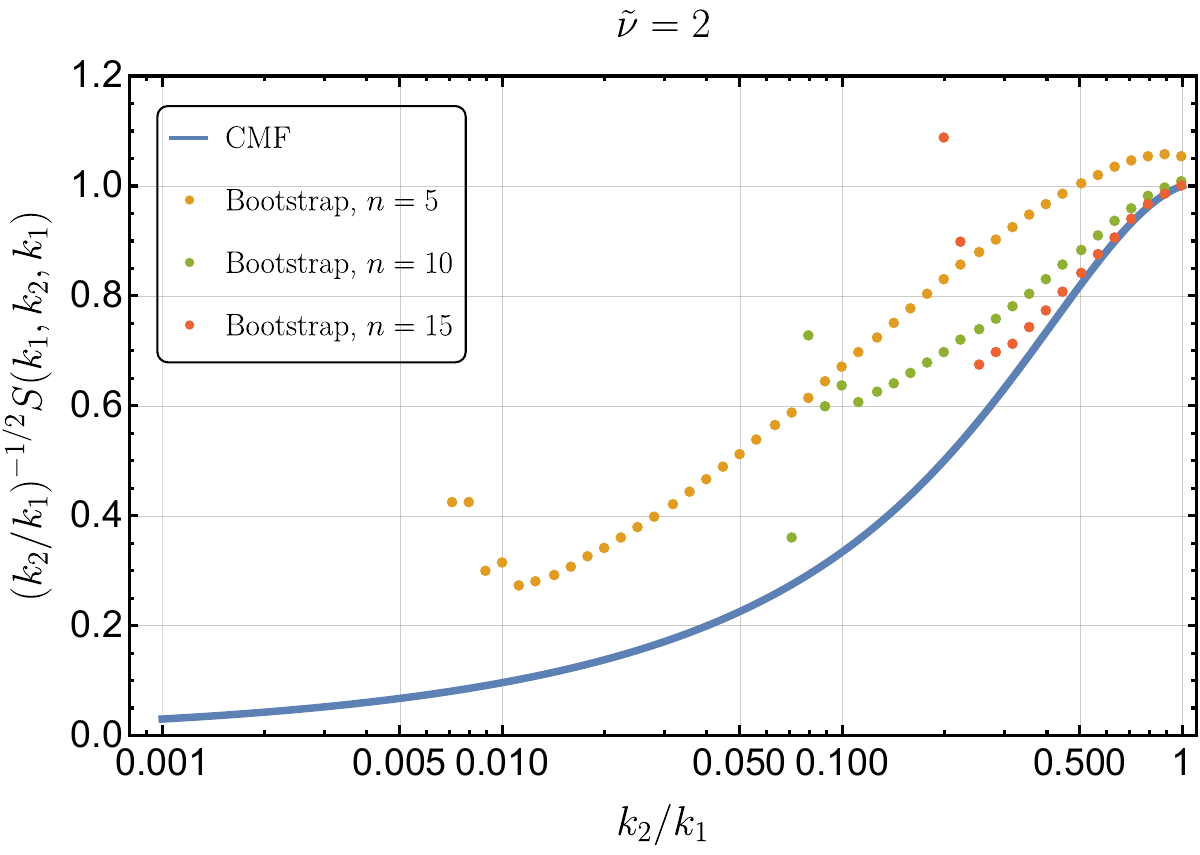}
    \caption{Comparison of CMF and bootstrap result for the permutation of double exchange diagram that does not contribute to oscillatory signals. In this figure we have plotted the bootstrap results summed to $(n+1)$th term for $n=5,10,15$. We find that the numerical instability in this case prevents us from reaching the result with an acceptable accuracy.}
    \label{fig:double_folded}
\end{figure}

Unlike the case of SE diagrams, the bootstrap formulae for the DE diagrams suffer a lot more from the numerical instability. As in the previous case, we evaluate the shape $S(k_1,k_2,k_1)$ produced by the middle diagram in Fig. \ref{fig:double_folded} for $\tilde{\nu}=2$ with both methods, and the result is shown in Fig. \ref{fig:double_folded}. Since the bootstrap formulae of the DE diagrams include power series which are not resummed, we need to decide the order of terms up to which we want to sum. In this figure, we show the result with different choices of cutoff order. It is clear to see that the result starts to lose numerical control at $k_2/k_1\sim0.01$ even if we cut off the series only at the 5th order, when the result is still very inaccurate as can be compared to the numerical result. As we increase the order of summation, although the result gradually converges to the numerical one, the numerical instability appears at even higher value of $k_2/k_1$. In conclusion, the bootstrap formulae do not work well in the whole momentum range region for observations, and numerical treatment is necessary for this case.

\begin{figure}[htbp]
    \centering
    \includegraphics[width=0.7\linewidth]{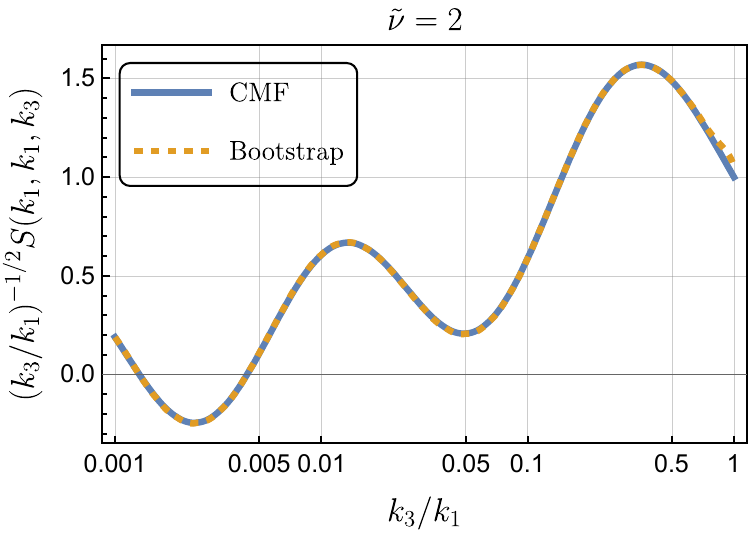}
    % \hspace{0.018\linewidth}
    \includegraphics[width=0.7\linewidth]{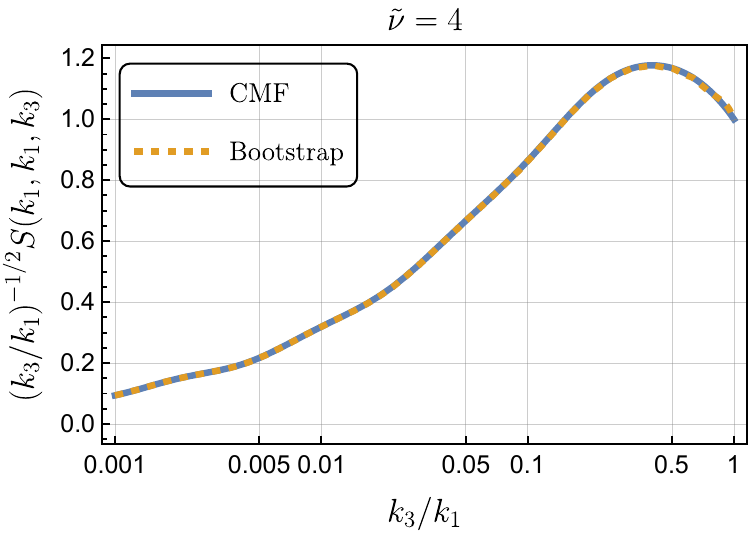}
    \includegraphics[width=0.7\linewidth]{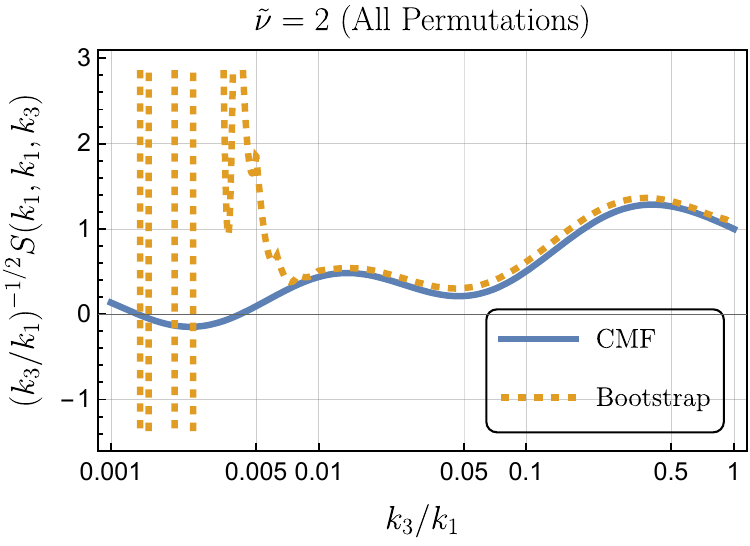}
    % \hspace{0.018\linewidth}
    \includegraphics[width=0.7\linewidth]{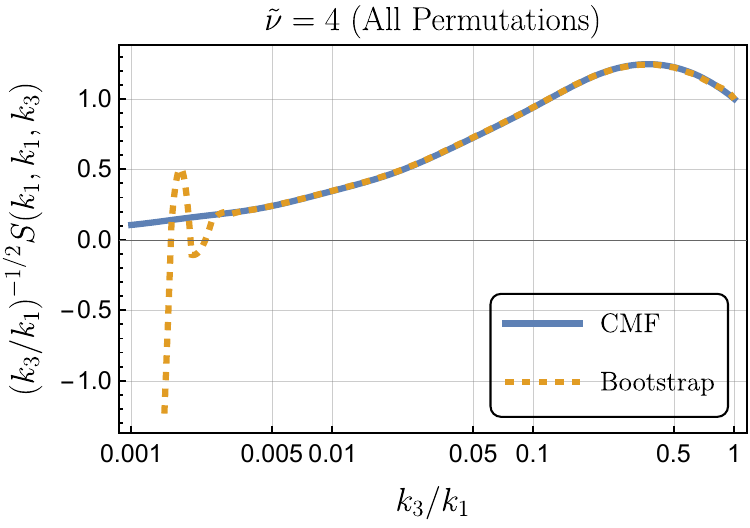}
    \caption{The shapes of double exchange diagrams for $\tilde{\nu}=2$ and $\tilde{\nu}=4$, evaluated by both two methods. The upper panels include only one diagram that contributes to oscillatory signals, and the lower panels include all permutations. In these figures the series in the bootstrap formulae are summed to the 5th order. It can be seen that although both methods work well for the diagram with signals, the numerical instability at the folded limit of the bootstrap formulae spoils the shapes when all permutations are added, while the CMF method is stable in the whole kinematic region.}
    \label{fig:double_shape}
\end{figure}

To see how severe the numerical instability can spoil the templates, we plot the shapes evaluated by both methods for two representative masses $\tilde{\nu}=2$ and $\tilde{\nu}=4$ in Fig.~\ref{fig:double_shape}. In this figure, we cut off the summation to the 5th order, which is a balance of accuracy and numerical control. Although we see from the upper panels that both methods work well for evaluating the oscillatory signals, the numerical instability spoils the templates after adding up all permutations. In particular, the problem is more severe for lower values of $\tilde{\nu}$ whose template are more interesting due to the large amplitude of oscillatory signals. Therefore, we need to use the CMF method to generate templates of the full shape.

\bibliography{references_letter}
\end{document}